\documentclass[preprint2]{aastex}
\usepackage{amssymb,amsmath}
\usepackage{graphicx}
\usepackage{longtable, float, subfigure}
\usepackage{geometry}
\usepackage{pdflscape, natbib}
\usepackage[T1]{fontenc}
\usepackage[latin1]{inputenc}
\def\kms{km s$^{-1}$}
\def\et{{et~al.}}
\def\ha{H$\alpha$}
\def\arcmin{\ifmmode^\prime\;\else$^\prime$\fi}
\def\arcsec{\ifmmode^{\prime\prime}\;\else$^{\prime\prime}$\fi}
\def\deg{\ifmmode^\circ\;\else$^\circ$\fi}

\def\hi{H\;\!{\sc i}}
\def\vmax{$V_{\rm max}$}
\def\sigz{$\sigma_{\rm z}$}
\def\msun{$M_{\sun}$}

\shorttitle{Stellar and Gas Kinematics of Dwarf Irregular Galaxies}
\shortauthors{Johnson}

\begin{document}
\title{The Shape of LITTLE THINGS Dwarf Galaxies DDO 46 and DDO 168: Understanding the stellar and gas kinematics}
\author {Megan C.\ Johnson\altaffilmark{1,2}
Deidre A.\ Hunter\altaffilmark{1}
Sarah Wood\altaffilmark{3,7}
Se-Heon Oh\altaffilmark{4,5}
Hong-Xin Zhang\altaffilmark{6}
Kimberly A.\ Herrmann\altaffilmark{1,8}
Stephen E.\ Levine\altaffilmark{1,9}
}

\altaffiltext{1}{Lowell Observatory, Flagstaff, AZ; dah@lowell.edu, sel@lowell.edu}
\altaffiltext{2}{CSIRO Astronomy and Space Science, Australia Telescope National Facility, Marsfield, NSW, Australia; megan.johnson@csiro.au}
\altaffiltext{3}{National Radio Astronomy Observatory, Green Bank, WV; swood@nrao.edu}
\altaffiltext{4}{International Centre for Radio Astronomy Research (ICRAR), University of Western Australia, Crawley, WA, Australia; se-heon.oh@uwa.edu.au}
\altaffiltext{5}{ARC Centre of Excellence for All-sky Astrophysics (CAASTRO)} 
\altaffiltext{6}{The Kavli Institute for Astronomy and Astrophysics, Peking University, Beijing, China; hongxin@pku.edu.cn}
\altaffiltext{7}{University of Tennessee, Knoxville, TN, USA}
\altaffiltext{8}{Pennsylvania State University, Mont Alto, PA, USA}
\altaffiltext{9}{Department of Earth, Atmospheric and Planetary Sciences, MIT, Cambridge, MA, USA}

\begin{abstract}

Determining the shape of dwarf irregular (dIrr) galaxies is controversial because if one assumes that these objects are disks and if these disks are randomly distributed over the sky, then their projected minor-to-major axis ratios should follow a particular statistical distribution, which is not observed.  Thus, different studies have led to different conclusions.  Some believe that the observed distributions can be explained by assuming the dIrrs are \emph {thick} disks while others have concluded that dIrrs are triaxial.  Fortunately, the central stellar velocity dispersion, $\sigma_{\rm z,0}$, combined with maximum rotation speed, \vmax, provides a kinematic measure, \vmax/$\sigma_{\rm z,0}$, which gives the three dimensional shape of a system.  
In this work, we present the stellar and gas kinematics of DDO 46 and DDO 168 from the LITTLE THINGS survey and determine their respective \vmax/$\sigma_{\rm z,0}$ values.  We used the Kitt Peak National Observatory's Mayall 4-meter telescope with the Echelle spectrograph as a long-slit spectrograph, which provided a two dimensional, 3$\arcmin$-long slit.  We acquired spectra of DDO 168 along four position angles by placing the slit over the morphological major and minor axes and two intermediate position angles.  However, due to poor weather conditions during our observing run for DDO 46, we were able to extract only one useful data point from the morphological major axis.  We determined a central stellar velocity dispersion perpendicular to the disk, $\sigma_{\rm z,0}$, of 13.5 $\pm$ 8 \kms\ for DDO 46 and <$\sigma_{\rm z,0}$> of 10.7 $\pm$ 2.9 \kms\ for DDO 168.  We then derived the maximum rotation speed in both galaxies using the LITTLE THINGS \hi\ data.  We separated bulk motions from non-circular motions using a double Gaussian decomposition technique and applied a tilted-ring model to the bulk velocity field.  We corrected the observed \hi\ rotation speeds for asymmetric drift and found a maximum velocity, \vmax, of 77.4 $\pm$ 3.7 and 67.4 $\pm$ 4.0 for DDO 46 and DDO 168, respectively.  Thus, we derived a kinematic measure, \vmax/$\sigma_{\rm z,0}$, of 5.7 $\pm$ 0.6 for DDO 46 and 6.3 $\pm$ 0.3 for DDO 168.  Comparing these values to ones determined for spiral galaxies, we find that DDO 46 and DDO 168 have \vmax/$\sigma_{\rm z,0}$ values indicative of \emph {thin} disks, which is in contrast to minor-to-major axis ratio studies.

\end{abstract}
\keywords{galaxies: individual (DDO 46, DDO 168) --- galaxies: dwarf galaxies --- galaxies: kinematics}

\section{Introduction}\label{sec:intro}

When classifying dwarf galaxies, it is commonly stated in the literature that gas rich dwarf irregular (dIrr) galaxies are the low luminosity tail of the spiral galaxy Hubble tuning fork classification sequence \citep[e.g.,][]{dev59, bin88}.  If this hypothesis is correct, then one would expect these objects to be intrinsically disk shaped like the more massive spiral galaxies.  If one assumes that dIrr galaxies are, in fact, disks and if those disks are randomly distributed across the sky, then the projected minor-to-major axis ratios (\emph{b/a}) and the corresponding derived inclination angles should follow a similar statistical distribution as for the spiral disk galaxies.  However, this is not what previous studies have shown and different groups have interpreted the inconclusive results to mean different things.  
\citet{sta92} and \citet{hod66} surmized that their observed \emph{b/a} statistical distributions implied that dIrr galaxies are thick disks.  
Others have interpreted the distribution as evidence of a triaxial nature \citep{bin95},
only a little less spherical than dwarf ellipticals \citep{sun98}.  Recent studies by \citet{san10} assume galaxies are oblate spheroids and find a trend in disk thickness with galaxy mass; dIrr galaxies are thinnest near stellar mass M$_*$ $\sim$ 2$\times$10$^9$ \msun\ and thicken on either side of this mass.  In contrast, \citet{roy13} made no assumption on the intrinsic shapes and found that dIrr disks thicken with decreasing luminosity; the most massive, brightest dIrr systems have the thinnest disks and the disks thicken as luminosity decreases.

In order to resolve the \emph{b/a} controversy and to determine if a disk thickness with luminosity trend does, in fact, exist, a more robust method for measuring the three-dimensional shape of dIrr systems is necessary.  The stellar kinematics, in particular, the central ($R=0$) stellar velocity dispersion perpendicular to the disk, $\sigma_{\rm z,0}$, combined with the maximum rotation speed, \vmax, usually determined from \hi\ kinematics, produces a kinematic measure, \vmax/$\sigma_{\rm z,0}$.  If an object is a disk, then it will be dominated by rotation and supported by angular momentum (\vmax\ $>$ \sigz) \citep{bin98}; on the other hand, a triaxial system is pressure supported and, therefore, dominated by the random motions of the stars (\vmax\ $<$ \sigz).  Thus, a spiral galaxy disk has \vmax/$\sigma_{\rm z,0}$\ $>$ 1, usually between two and five \citep{bot93, veg01, bec04, mar13}, and triaxial systems such as giant elliptical and dwarf elliptical (dE) galaxies have \vmax/$\sigma_{\rm z,0}$ $\le$ 1 \citep{bin82, bin93, ped02}.  Thus, \vmax/$\sigma_{\rm z,0}$ is a definitive method for understanding the intrinsic three-dimensional shape of a galaxy. 

There is evidence in the literature that dIrr galaxies may be an earlier evolutionary stage of dSph and dE systems.  Through ram pressure stripping, harassment, tidal forces, and other interaction mayhem, dIrr galaxies may morph into dE, like the ones found in the Virgo cluster that contain embedded stellar disks \citep{lis07}, or they may become dSph systems like those observed around the Milky Way and M31 \citep[e.g.][]{gre03}.  Thus, dIrr galaxies may be innately triaxial, or, through interaction processes, dIrr systems may transform from a rotationally supported disk to a triaxial pressure supported system.  By studying the stellar and gas kinematics of nearby dIrr galaxies, we may get a glimpse of the potential mechanism(s) for transforming a dIrr system into dE or dSph systems. 

\begin{deluxetable}{lccc}
\tabletypesize{\scriptsize}
\tablenum{1}
\tablecolumns{4}
\tablewidth{0pt}
\tablecaption{Global Parameters for DDO 46 and DDO 168}
\tablehead{
\colhead{Parameter} & \colhead{DDO 46} &\colhead{DDO 168} &\colhead{Ref}
}
\startdata
Other Names & UGC 03966, PGC 21585 
& UGC 08320, PGC 46039 & 1\\
Distance (Mpc) & 6.1 
& 4.3 & 2\\
M$_V$ & $-$14.7 
&$-$15.7 &2\\
Galaxy diameter to 25 mag arcsec$^{-2}$ in $B$, D$_{25}$ (arcmin) &1.49 
&2.85& 3\\
$V$-band disk scale length, R$_D$ (kpc) & 1.14 $\pm$ 0.06 
&0.82 $\pm$ 0.01 & 2\\
Center (RA, DEC) (J2000) & (07:41:26.6, $+$40:06:39)
&(13:14:27.2, $+$45:55:46) & 4\\
Minor-to-major axis ratio, $b/a$ &0.89
&0.63 & 4\\
Inclination from $b/a$, $i_{opt}$ (degrees) & 30  
& 58 & \\
L$_{H\alpha}$ (ergs s$^{-1}$) & (6.17 $\pm$ 0.14) $\times$ 10$^{38}$ 
&1.07 $\times$ 10$^{39}$& 3\\
log Star Formation Rate (log SFR$_D$) (M$_{\odot}$yr$^{-1}$kpc$^{-2}$) & $-$2.89 $\pm$ 0.01
&$-$2.27 $\pm$ 0.01 &2\\

\cutinhead{Stellar parameters determined from this work}
$V_{\rm sys}$ (km s$^{-1}$) & \nodata &185.9 $\pm$ 9.9 & \\
Kinematic Major Axis PA (degrees) & \nodata  &288& \\
Average Central Velocity Dispersion, $<\sigma_{\rm z,0}>$ (\kms) & 13.5 $\pm$ 8 
&10.7 $\pm$ 2.9&\\
Total Stellar Mass -- 3.6$\mu$m (log $M_\sun$) & \nodata &7.73&\\
Total Stellar Mass -- SED (log $M_\sun$) & 7.39$^{+0.16}_{-0.09}$& 7.77$^{+0.15}_{-0.16}$&\\
Optical scale-height, $h_{\rm z}$ (pc)&155&110&\\

\cutinhead{H\;\!{\sc i} parameters determined from this work}
$V_{\rm sys}$ (km s$^{-1}$),  & 360.8 $\pm$ 1.3 &192.6 $\pm$ 1.2& \\
Kinematic Major Axis PA (degrees) & 274.1 $\pm$ 5 &                275.5 $\pm$ 5.8 & \\
Inclination from tilted ring model, $i_{HI}$ (degrees) &27.9 $\pm$ 0.1 &                   46.5 $\pm$ 0.1 &\\
Kinematic Center ($X_{\rm pos}$, $Y_{\rm pos}$) (RA, DEC)  &(07:41:26.3, +40:06:37.5) &           (13:14:27.3, +45:55:37.3) & \\
$V_{\rm rot}$ (\kms) &69.7 $\pm$ 0.6 &             54.7 $\pm$ 0.3&\\
$V_{\rm max}$ (\kms) & 77.4 $\pm$ 3.7&             67.4 $\pm$ 4.0&\\
Average Velocity Dispersion, $<\sigma_{\rm HI}>$ (\kms) &6.5 &                 11.0&\\
Central Velocity Dispersion, $\sigma_{\rm HI, 0}$ (\kms) &10.0 &                 12.0&\\
Total Gas Mass ($M_\sun$) &2.2 x 10$^8$ & 2.6 x 10$^8$&\\
$M_{\rm dyn}$ ($M_\sun$) & 4.1 x 10$^9$& 3.3 x 10$^9$&\\
&($R_{\rm max}$ = 2.92 kpc)&($R_{\rm max}$ = 3.14 kpc) &\\
$M_{\rm DM}$ ($M_\sun$)  & 3.8 x 10$^9$ & 3.0 x 10$^9$&\\
$V_{\rm max}$/$\sigma_{\rm z,0}$ &5.7 $\pm$ 0.6 & 6.3 $\pm$ 0.3 &\\
$M_{\rm HI}$/$M_{\rm *}$ & 8.9 & 4.4&\\
$M_{\rm DM}$/$M_{\rm bar}$ & 16 & 9&

\enddata
\tablerefs{(1) NASA Extragalactic Database; 
(2) \citet[][and references therein]{hun12}; 
(3) \citet{hun04}; 
(4) \citet{hun06} 
}\label{tab:1}
\end{deluxetable}

In this work, we present the stellar and \hi\ gas kinematics for DDO 46 and DDO 168, two nearby dIrr galaxies.  Both DDO 46 and DDO 168 are ``typical'' dIrrs representative of the LITTLE THINGS sample; they have absolute $V$ magnitudes, $M_{\rm V}$, of $-$14.7 and $-$15.7, respectively, which places them at the peak of the histogram distribution for magnitude versus number of galaxies for the dIrrs in LITTLE THINGS as shown in Figure 1 of \citet{hun12}.  Both have typical star formation rates, \hi\ masses, and disk scale lengths \citep{hun06, hun12}.

Table \ref{tab:1} lists some of the global parameters for our targets.
The stellar kinematics for our galaxies were obtained using the Kitt Peak National Observatory's (KPNO\footnote{The Kitt Peak National Observatory is operated by the National Optical Astronomy Observatory (NOAO), which is operated by the Association of Universities for Research in Astronomy (AURA) under cooperative agreement with the National Science Foundation.}) Mayall 4-meter telescope with the Echelle spectrograph. We obtained long-slit, two-dimensional optical spectra centered on the Mg Ib triplet along four position angles (PAs) per galaxy: major axis, minor axis, and $\pm$ 45$\arcdeg$ to the major axis.

The \hi\ gas kinematics are from high spatial and spectral resolution data obtained with the National Radio Astronomy Observatory's (NRAO\footnote{The National Radio Astronomy Observatory is a facility of the National Science Foundation operated under cooperative agreement by Associated Universities, Inc.}) Karl G. Jansky Very Large Array (VLA).  The data for DDO 46 and DDO 168 are from LITTLE THINGS (Local Irregulars That Trace Luminosity Extremes; The \hi\ Nearby Galaxy Survey) \citep{hun12}.

This paper is organized as follows: Section \ref{sec:optspec} describes the observations, data reductions, and analysis for the optical spectra; Section \ref{sec:hispec} describes the data and analysis for the \hi\ spectroscopy; Section \ref{sec:results} explores the results; Section \ref{sec:discuss} is our discussion section; Section \ref{sec:sumcon} presents the summary and outlines the conclusions.

\section{Optical Stellar Spectra}\label{sec:optspec}

	\subsection{Observations}\label{sec:optobs}

\begin{figure*}
\begin{center}
\subfigure[DDO 46]{\includegraphics[scale=.25]{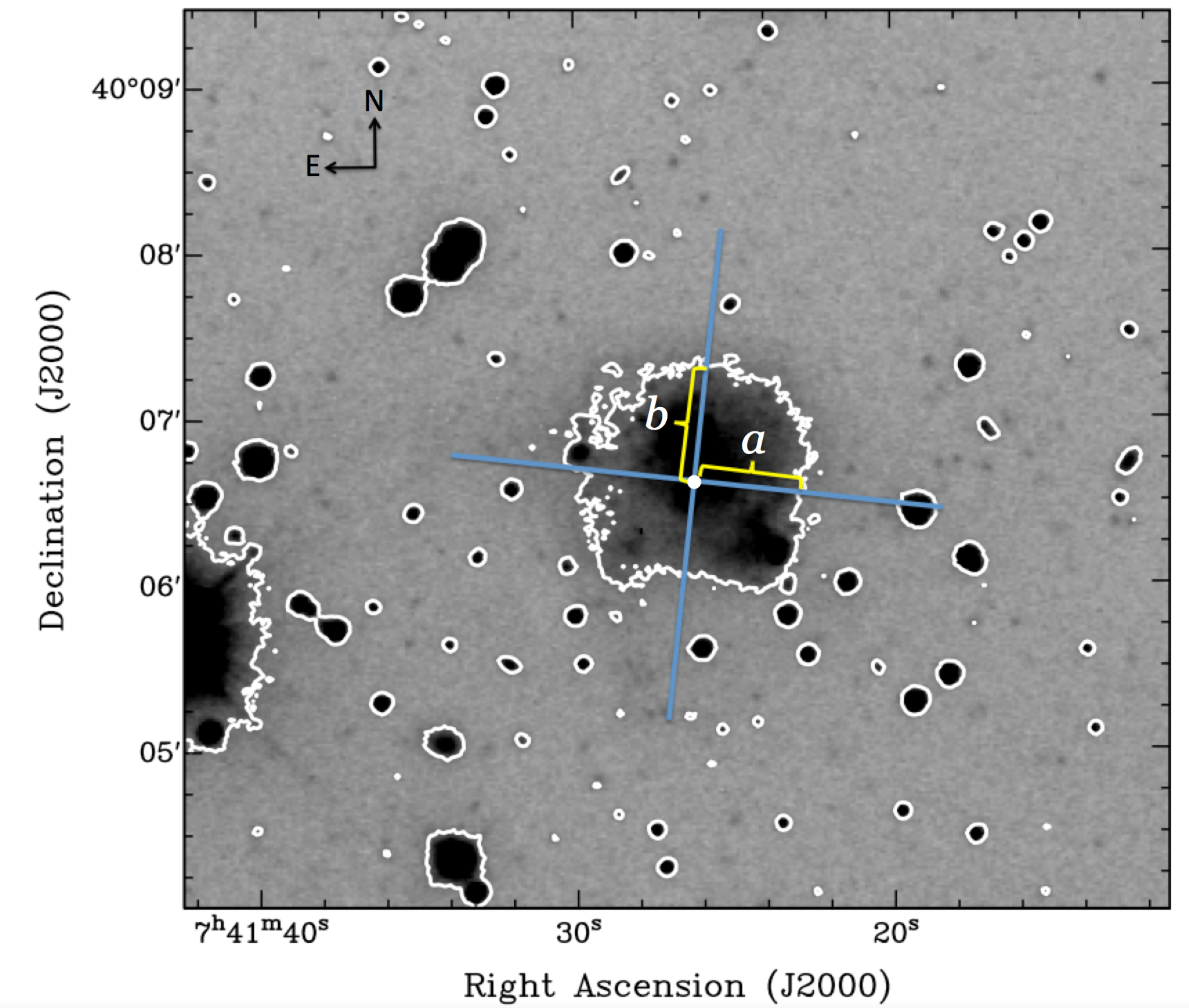}}\label{subfig:1a}
\subfigure[DDO 168]{\includegraphics[scale=.2]{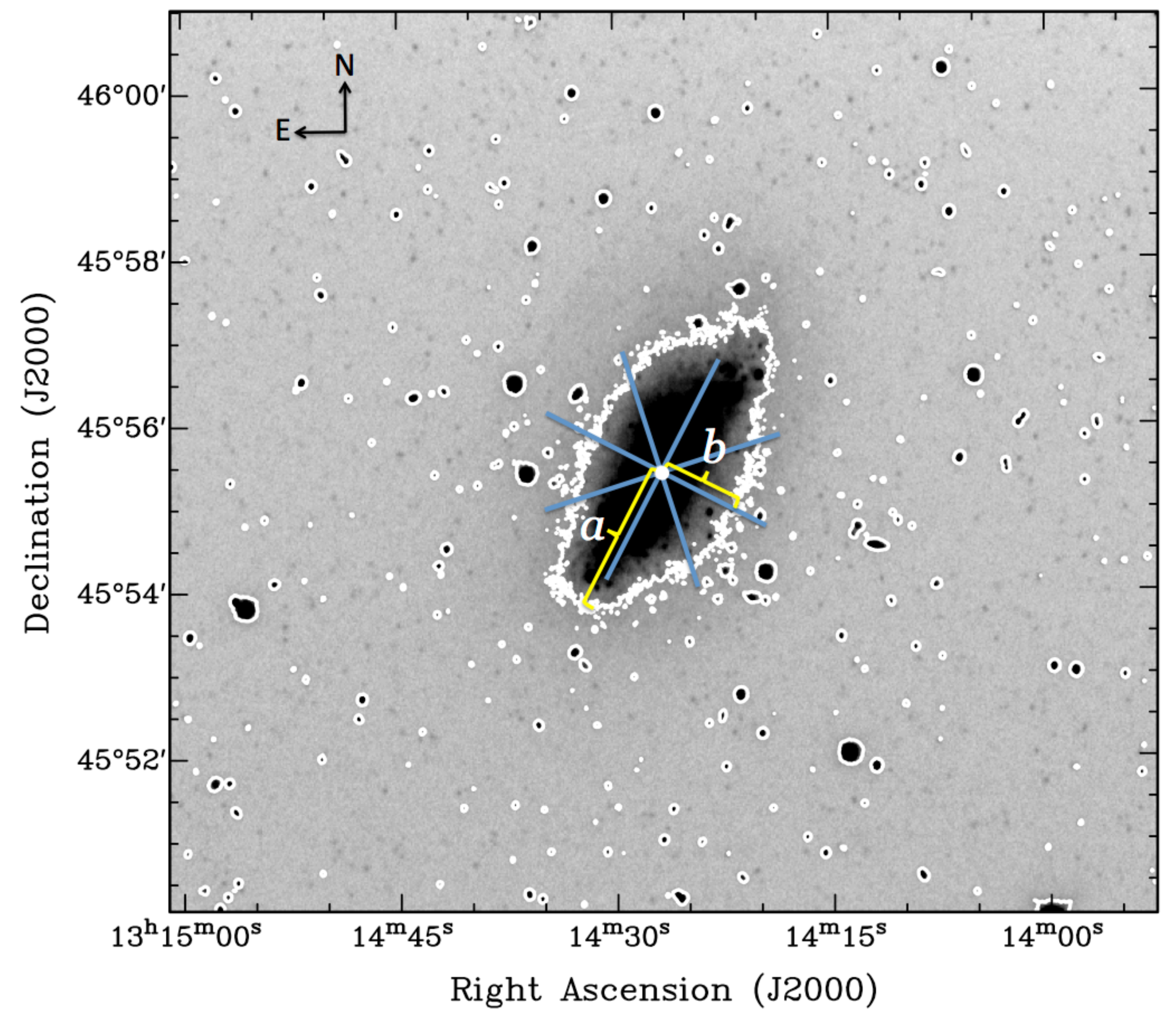}}\label{subfig:1c}
\caption{LITTLE THINGS $V-$band images of DDO 46 (a) and DDO 168 (b) with outermost contour used to determine morphological major and minor axes. The KPNO 4-meter $+$ Echelle spectrograph 3$\arcmin$ slit PAs are over plotted for each galaxy. The white dots mark the morphological center in each and the ``a'' and ``b'' letters mark the semi-major and semi-minor axes, respectively.  North is up and east is to the left.}
\label{fig:1}
\end{center}
\end{figure*}

Figure \ref{fig:1} shows $V-$band images for DDO 46 and DDO 168 from the LITTLE THINGS dataset
 with the 3$\arcmin$ slit positions over plotted.  The white dot in each image marks the morphological center and the center of the slit.  Figure \ref{fig:1} also includes the outermost $V-$band intensity contour, which is what is used to identify the morphological semi-major ($a$) and semi-minor ($b$) axes for each object.  Both $a$ and $b$ are marked on Figure \ref{fig:1} for each galaxy and our $b/a$ values and corresponding inclinations, $i$ (assuming an intrinsic disk thickness for irregulars, $q_{\rm o}$ = 0.4 \citep{van88}) agree with those from \citet{hun06}.  See Table \ref{tab:1} for the respective \emph{b/a} and $i$ values.

Both galaxies were observed with the KPNO Mayall 4-meter telescope and Echelle spectrograph. DDO 46 was scheduled for six nights in 2010 January along the major and minor morphological axes, but due to rainy, cloudy weather, only one night has usable data. 
DDO 168 was observed for two nights in 2008 February along the major axis and then again in 2009 April and it was observed in all four PAs, morphological major and minor axes and intermediate angles $\pm$45$\arcdeg$ from the major axis.  The weather for all of the nights we observed DDO 168 was superb, with the exception of 3 Feb 2008 where some clouds reduced the data quality.  

Table \ref{tab:2} shows a summary of the observations.  We used the 4-meter$+$Echelle spectrograph as a 3$\arcmin$ long-slit, two-dimensional spectrograph by replacing the cross-disperser with a mirror flat.  We opened the slit to 2$\farcs5$ in order to maximize throughput while maintaining high spectral resolution.  The average seeing at the KPNO 4-meter telescope during our observing run in April 2009 for DDO 168 was quite good and generally less than 2$\farcs5$, although there was one night when the seeing worsened and may have been about this.  The cloudy, rainy weather conditions that plagued our observing run in January 2010 for DDO 46 made for poor seeing conditions and during some nights caused seeing to jump above 2$\farcs5$. 
Thus, the point spread function (PSF) of the template stars
in the case of DDO 46 and on one night with DDO 168
were comparable to the slit-width and, thus, the spectral FWHM of the stars
were observed under comparable slit-illumination conditions as the galaxy.
However, on the other DDO 168 nights
the spectral FWHM of the stars was set by the seeing.
This may potentially lead to an overestimate
of the line-of-sight velocity dispersions
obtained for the galaxies. However, this effect
is likely minimal because 
a comparison of the FWHM of the stars on nights of good seeing are $<$ 15\% 
different compared to the FWHM of stars
observed under poorer seeing conditions.

We targeted the strong Mg Ib stellar absorption features at 5167.3 \AA, 5172.7 \AA, and 5183.6 \AA\ by isolating a single order using a filter with a central wavelength of 5204 \AA\ and full-width at half-maximum (FWHM) of 276 \AA.  These Mg Ib absorption features are easily detected because they are deep and this wavelength region is virtually free of telluric features. 
Our setup produces a FWHM $=$ 0.38 \AA\ as measured from the central lines of Th-Ar comparison lamp exposures and gives a velocity resolution of $\sim$22 \kms, which corresponds to a velocity dispersion, $\sigma$, of $\sim$9 \kms.  We acquired an angular pixel scale of 1$\farcs2$ pix$^{-1}$, after binning by two pixels in the cross-dispersion direction, and a spectral pixel scale of 0.14 \AA\ pix$^{-1}$.   

On the first night of every observing run, we stepped a radial velocity standard star in equal spatial steps along the cross-dispersion direction of the slit in order to model the cross-dispersion curvature and to obtain the spatial pixel scale.  We observed several radial velocity standard stars at the beginning and ending of every night and acquired comparison lamp exposures at each star location in order to accurately solve for the wavelength solution.  Table \ref{tab:3} lists all of the radial velocity standard stars observed each night along with their exposure times, spectral types and classes, magnitudes, coordinates, metallicities (where available), and heliocentric radial velocities.  We used these radial velocity standard stars in the Milky Way as templates and applied a cross-correlation method (CCM) to determine the velocities and intrinsic velocity dispersions of the stars in each galaxy.  

Careful sky subtraction is essential to our analysis and this was achieved by observing equal amounts of time on the galaxy as blank sky near the target position.  
Each night, we observed one PA of the galaxy and took several 1800 s exposures for each galaxy position and several 1800 s exposures of blank sky near the galaxy position in order to properly subtract sky.  Comparison lamp exposures were taken before and after each galaxy and sky exposure so that accurate wavelength calibration could be performed. 

		\subsection{Data Reductions}\label{sec:optred}
		
All optical stellar spectra were reduced using the Image Reduction and Analysis Facility (IRAF\footnote{IRAF is distributed by the National Optical Astronomy Observatory.}).  We presented the detailed data reduction process in a companion paper on NGC 1569 and refer the reader to \citet{joh12} for more comprehensive information.  Thus, we only briefly outline the major steps taken in the data reduction process here.  

We began our data reductions by fitting for the overscan to remove the electronic pedestal in each image. Then, we averaged and normalized 10 dome flats taken each night to correct our spectra for any pixel variations or imperfections from the instruments. We averaged and normalized a series of twilight sky flats 
to correct for illumination variations along the cross-dispersion direction of the slit.  We took bias frames at the start of each observing run to ensure that there was no electronic noise pattern across the CCD but because no significant variations were observed, we did not use the bias exposures in our data reductions.

We used the radial velocity standard star that was moved in equal spatial steps along the slit to correct for the slit curvature.  We used comparison lamp exposures to find our wavelength solutions and found that our spectra are accurate to 0.4 \AA\ for a given night, which enabled us to put all spectra for a night on a single wavelength scale.  We co-added all of the galaxy exposures together to maximize signal-to-noise (S/N) and then subtracted the co-added sky exposures to produce a single, two-dimensional spectrum.  
Table \ref{tab:2} lists the total integration times for each galaxy and sky position per night.

%

\begin{deluxetable}{rcccccccc}
\tabletypesize{\small}
\tablenum{2}
\tablecolumns{8}
\tablewidth{0pt}
\tablecaption{KPNO 4-m + Echelle spectroscopic observations}
\tablehead{\colhead{UT Date} & \colhead{PA\tablenotemark{a}} & \colhead{$N_{\rm gal}$\tablenotemark{b}} & \colhead{$T_{\rm gal}$(s)\tablenotemark{c}} & \colhead{$N_{\rm sky}$\tablenotemark{d}} & \colhead{$T_{\rm sky}$(s)\tablenotemark{e}} & \colhead{$N_{\rm spec}$\tablenotemark{f}} & \colhead{$N_{\rm RV}$\tablenotemark{g}} & \colhead{Weather}\\
\colhead{Obs.}&&&&&&&&\colhead{Conditions}}
\startdata
\cutinhead{DDO 46}
15 Jan 10 &84$\arcdeg$ (major axis) & 5 & 9000 & 5 & 9000 & 0& 4& Clouds, rain\\
16 Jan 10 &84$\arcdeg$(major axis)&5&9000&5&7393&0&5 $+$ 1& Clouds, rain\\
17 Jan 10 &84$\arcdeg$(major axis) &4&7200&4&5673&1&11 $+$ 1&Strong winds, clear\\
18 Jan 10 &174$\arcdeg$(minor axis)&8&13,800&7&12,000&0&5&Heavy clouds\\
\cutinhead{DDO 168}
02 Feb 08&153$\arcdeg$(major axis)&4&7200&4&6900& 3&5$+$1&Clear\\
03 Feb 08&153$\arcdeg$(major axis)&4&7200&4&7200& 2&12$+$1&Clear, partly cloudy\\
17 Apr 09&63$\arcdeg$(minor axis)&6&10800&5&9000& 2&6&Clear\\
18 Apr 09&153$\arcdeg$(major axis)&6&10800&6&10800& 3&6$+$1&Clear\\
19 Apr 09&198$\arcdeg$&7&12600&7&12600& 2&8$+$1&Clear\\
20 Apr 09&108$\arcdeg$&7&12600&7&12600& 3&11$+$1&Clear\\
\enddata
\tablenotetext{a}{Position angle observed, parenthesis identify morphological position angle;}
\tablenotetext{b}{Number of galaxy exposures co-added into a single spectrum;}
\tablenotetext{c}{Total integration time, in seconds, on galaxy;}
\tablenotetext{d}{Number of sky exposures co-added into a single spectrum;}
\tablenotetext{e}{Total integration time, in seconds, on sky;}
\tablenotetext{f}{Number of one-dimensional spectra extracted;}
\tablenotetext{g}{Number of radial velocity standard stars used in cross-correlation method.  The ``$+$1'' is in reference to the solar spectrum, from twilight flats, used as a radial velocity standard star. }\label{tab:2}
\end{deluxetable}

			\subsection{Extraction of one-dimensional spectra}\label{sec:onespec}

We summed along lines parallel to the cross-dispersion direction in order to extract one-dimensional spectra.  This spatial summing process was necessary to produce adequate S/N in all of the one-dimensional spectra.  To preserve the spatial resolution of the data for comparing the stellar and gas kinematics at a given location, we limited the spatial area summed to less than approximately three \hi\ beam widths (see Section \ref{sec:hidata}).  For DDO 168, we summed 24$\farcs6$ (513 pc at a distance of 4.3 Mpc) to produce one-dimensional spectra along the morphological major axis and the two intermediate PAs.  For the data taken along the morphological minor axis, we summed over 12$\farcs3$, or 256 pc, because of the increase in S/N along this PA.  For DDO 46, we summed over 12$\farcs3$ (364 pc at a distance of 6.1 Mpc).  After the one-dimensional galaxy spectra were extracted, we fit the continuum.

We also extracted one-dimensional spectra from the radial velocity standard stars and fit for the continuum in each. We used these stellar templates to create our own template library and only the stars taken on the same night were used in the CCM.  This was to ensure that we did not introduce any potential night to night variations.  On the nights that we observed twilight flats, we extracted a solar spectrum for use as an additional template star in the CCM.

\begin{figure*}
\centering
\subfigure[DDO 46]{\includegraphics[scale=.25]{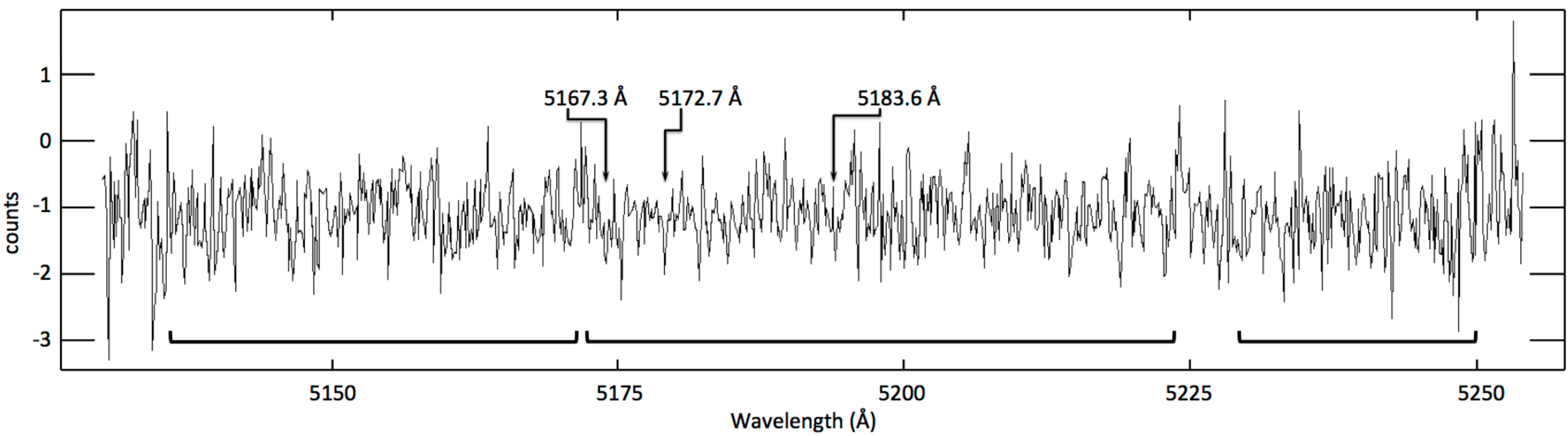}\label{subfig:2a}}
\subfigure[DDO 168]{\includegraphics[scale=.25]{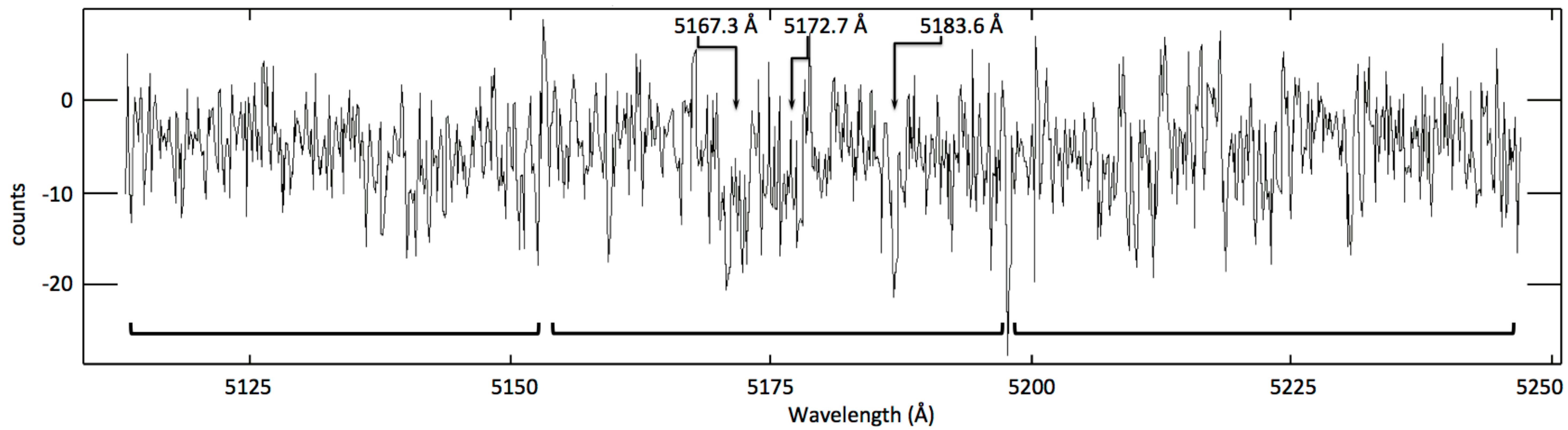}\label{subfig:2c}}
\subfigure[HD 132737]{\includegraphics[scale=.25]{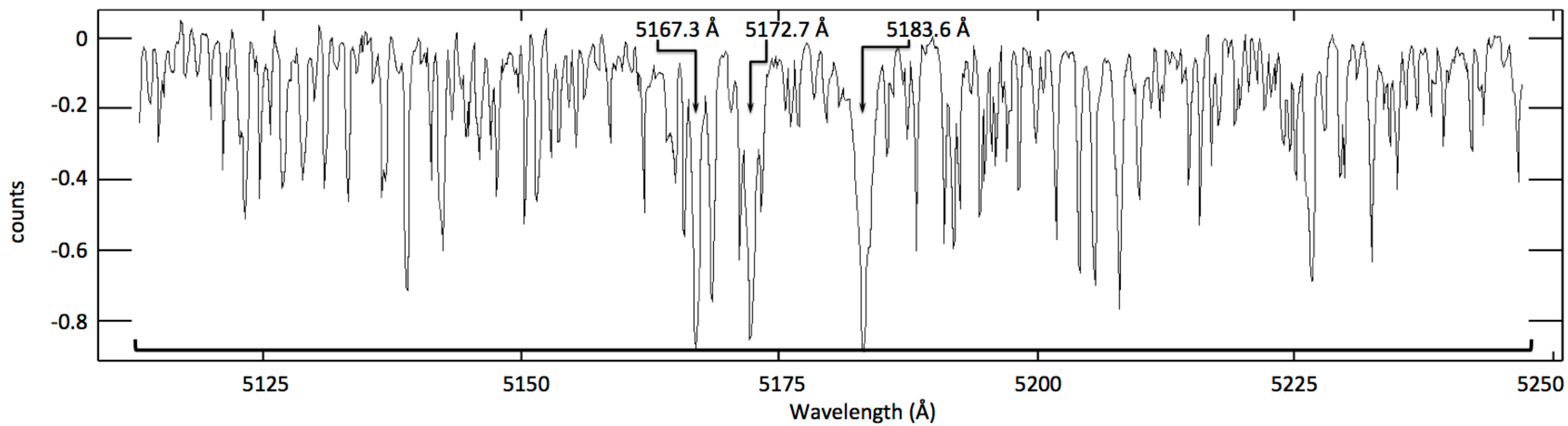}\label{subfig:2d}}
\caption{Example of continuum subtracted, one-dimensional spectra for the major axes of DDO 46 (a), DDO 168 (b) and one template stellar spectrum of HD 132737 (c) for comparison.   
The spectrum for DDO 46 has a S/N of $\sim$3 and was extracted at a radius of $-4.8\arcsec$(west) from center. 
The spectrum for DDO 168 has S/N of $\sim$6 and was extracted at a distance of +26$\farcs8$(east) from the center.  The stellar template spectra is of HD 132737 and has a S/N of $\sim$40, which was typical throughout the observing runs.
We used the regions shown by the horizontal brackets in the cross-correlation method (CCM). The rest wavelengths of the Mg Ib absorption triplet features are marked in each panel.}
\label{fig:2}
\end{figure*}

Figure \ref{fig:2} shows examples of continuum subtracted, one-dimensional galaxy spectra along the major axes for DDO 46, DDO 168 and a stellar template for comparison.  The rest wavelengths of the Mg Ib triplet absorption features are marked for reference and the regions used in the CCM are shown by the black horizontal brackets.  
 	
	\subsection{Analysis}\label{sec:optan}          
		
We used the IRAF task {\sc fxcor} \citep{fit93, ton79} to cross-correlate the galaxy spectra with the stellar template library from the same 
night.  
Proper sky subtraction is critical to produce a high S/N cross-correlation peak.
Thus, we tested all of the available scaling functions in IRAF in the co-adding process before, during, and after sky subtraction and analyzed the noise statistics and strength of the peak in the cross-correlation function to find the best sky-subtracted, extracted spectra.  We explored the effects of various spectral regions used in the cross-correlation function making sure to avoid any residual sky lines.  The main outcome from all of our tests is that the heliocentric radial velocities of the stars in the galaxy, be it DDO 46 or DDO 168, as determined from {\sc fxcor}, did not change by more than 10\% regardless of which scaling function (or none at all) or spectral regions we used. We did, however, achieve a higher signal-to-noise cross-correlation peak depending on the quality of sky subtraction and avoidance of residual sky lines, which in turn, produced a more robust (smaller standard deviation) stellar velocity dispersion. Because the data quality was so good for DDO 168, the sky subtraction issues didn't cause much variance in the resulting velocity dispersions.  However, because of the poor weather conditions and troubles with sky subtraction, there was only one extracted spectrum from DDO 46 that produced a cross-correlation peak above 3$\sigma$.

Table \ref{tab:3} shows the dates that the standard stars were observed and all stars observed on the same night make up the number of stars used in the CCM.  For example, to obtain the stellar kinematics for the morphological major axis of DDO 168 observed on 18 April 2009, we used the following seven stellar templates also observed on 18  April 2009 in the CCM: HD 66141, HD 92588, HD 132737, HD 145001, HD 182572, HD 187691, and the Sun.  We derive the heliocentric radial velocities of the stars in each object using the peaks of the CCM.  

\begin{figure*}
\centering
\subfigure[DDO 46]{\includegraphics[scale=.5]{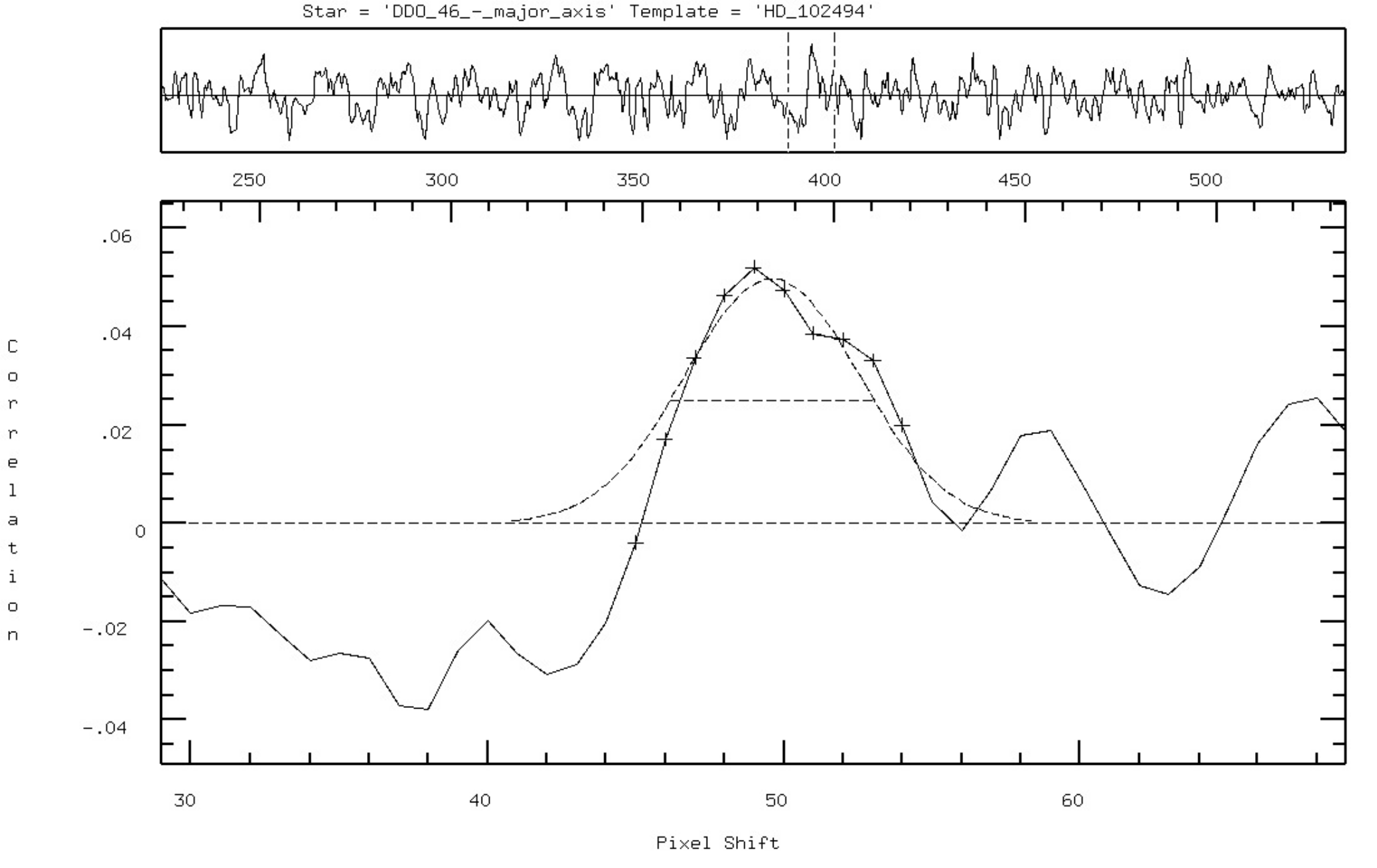}\label{subfig:3a}}
\subfigure[DDO 168]{\includegraphics[scale=.23]{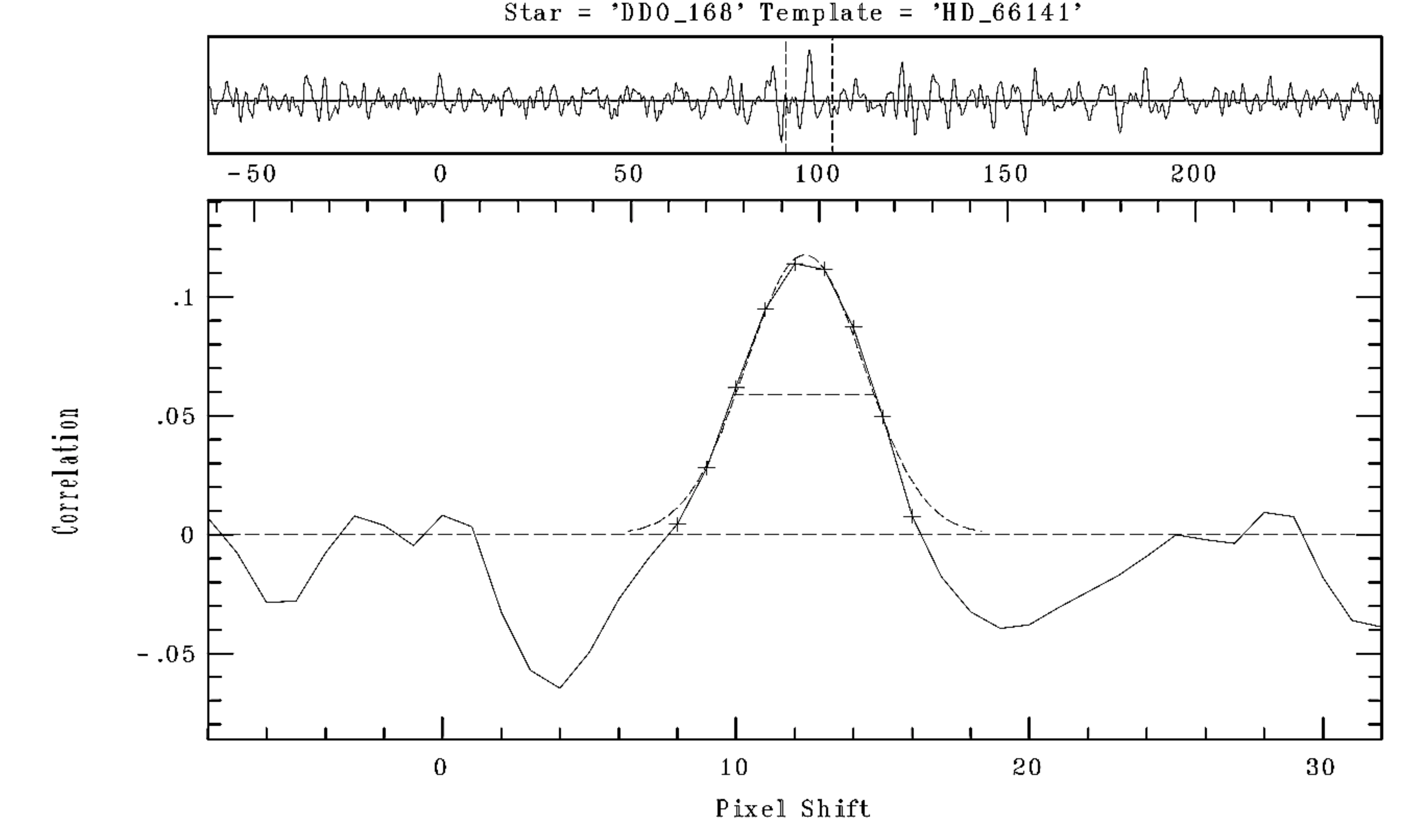}\label{subfig:3c}}
\caption{Example of cross-correlation peaks for DDO 46 (a) and DDO 168 (b) cross-correlated with radial velocity standard star, HD 102494 and HD 66141, respectively.}
\label{fig:3}
\end{figure*}

For each night, we cross-correlate the standard star templates with themselves in order to check the radial velocities and also to model the instrumental FWHM ($FWHM_{\rm instr}$) of the cross-correlation peak.  On average, we found that the $FWHM_{\rm instr}$ $\sim$ 30 \kms.  We subtracted the $FWHM_{\rm instr}$ in quadrature from the FWHM of the individual galaxy spectra cross-correlated with the template stars ($FWHM_{\rm obs}$) in order to determine the intrinsic velocity dispersion of the galaxies.  In equation \ref{eqn:sigma}, we use the definition of a Gaussian to derive the \emph {observed} stellar velocity dispersions of the stars:
\begin{equation}\label{eqn:sigma}
\sigma_{\rm obs} = \frac{\sqrt{{\rm FWHM_{obs}}^2 - {\rm FWHM_{instr}}^2}}{{2.35}} 
\end{equation}
Figure \ref{fig:3} shows example cross-correlation peaks for the cross-correlation between the summed regions of each galaxy and a single radial velocity standard star.  The S/N ratio for the cross-correlation peak is above 3 for each spectrum and because of the low S/N in the peak of DDO 46, the profile has a bumpy appearance.  

We observed as many different spectral types and classes (and metallicities, when available) of radial velocity standard stars each night (of which there aren't that many to choose from in a given night) in order to create as diverse a stellar template library as possible.   The spectrum of HD 132737 in Figure \ref{fig:2} is representative of how the stellar template spectra have deeper, more well defined absorption features than the galaxy spectra.  It is difficult to measure the effect of this template mismatch on the velocity dispersion results.  However, the FWHM values determined from the CCM for DDO 46 and DDO 168 are all remarkably consistent for a given galaxy spectrum and do not vary by more than 10\% between the template$-$galaxy pairs. 
Also, we found that there is less than a 10\% difference between the FWHM values from the cross-correlations between the different pairs of template stars, which is the same as for the template$-$galaxy pairs.  We noticed that the solar spectrum has consistently higher FWHM values when cross-correlated with the other templates (although only slightly and all within 10\% of the other templates) and this is likely attributed to the Sun's higher log(g) value compared to the giant template stars. We determined the errors on the instrumental dispersion  by determining the standard deviation of the FWHM values from {\sc fxcor} for each template star using the mean of the template$-$template pairs for each template and then, using the standard deviations as weights, we calculated the weighted average of all the templates from a given night to derive the final instrumental FWHM.
 Thus, our errors are reflective of any systematic effects.

The results of the CCM provide the $observed$ stellar velocity dispersions along the line of sight.  In order to determine if DDO 46 and DDO 168 are disks or triaxial in nature, we require the central velocity dispersion perpendicular to the disk, $\sigma_{\rm z,0}$, in order to derive \vmax/$\sigma_{\rm z,0}$.  Therefore, we need to decompose $\sigma_{\rm obs}$ into its three velocity components, $\sigma_{\rm R}$, $\sigma_{\rm z}$, and $\sigma_{\rm \phi}$. 

\begin{figure*}
\centering
\includegraphics[scale=.5]{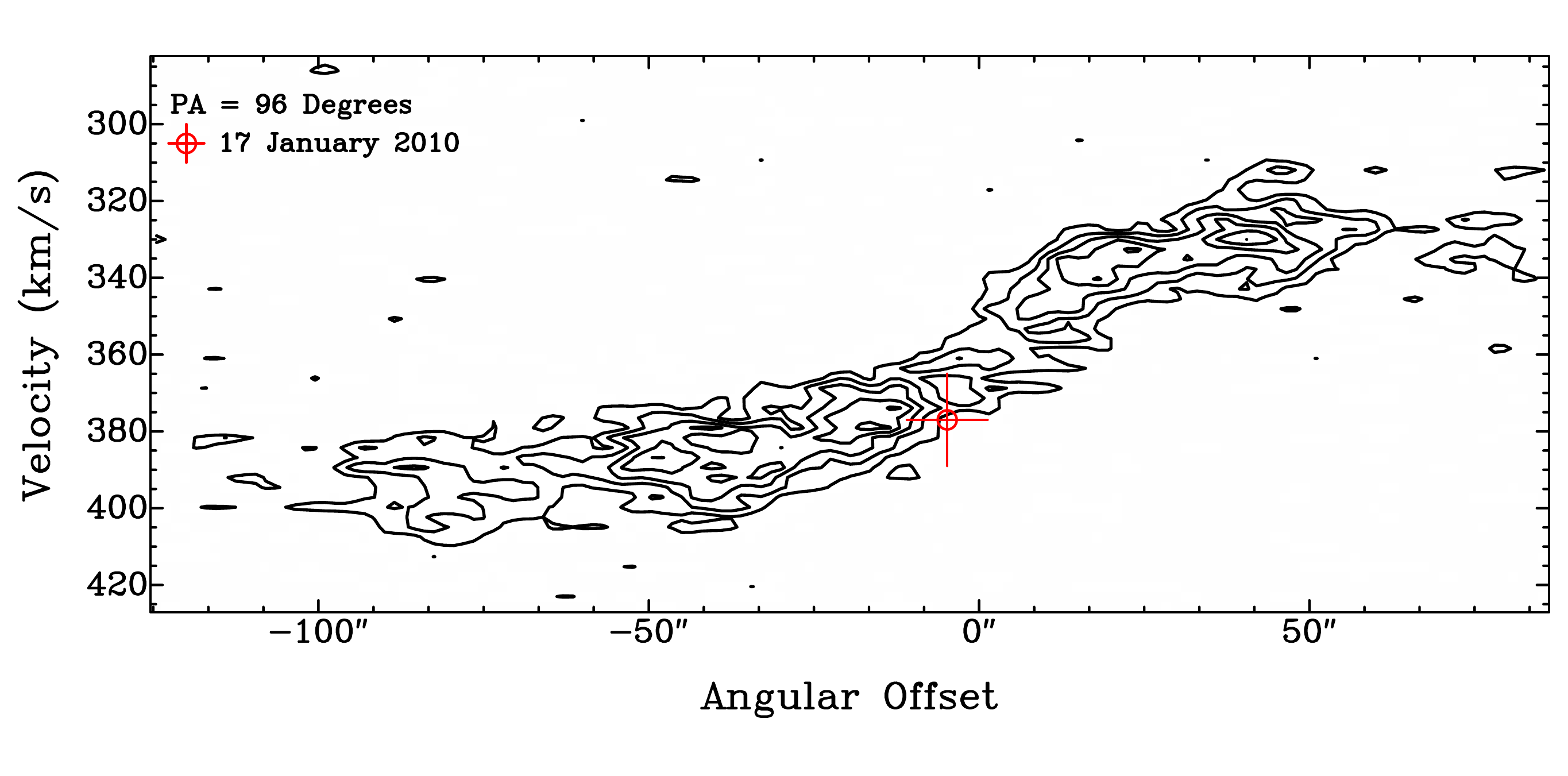}\label{subfig:4}
\caption{Position-velocity (P-V) diagram showing morphological major axis, PA = 96$\arcdeg$, along which we determined the stellar kinematics for DDO 46. The \hi\ is shown as contours and stellar velocity data point is plotted as a red plus sign for DDO 46.}
\label{fig:4}
\end{figure*}

\begin{figure*}
\centering
\subfigure[PA = 153$\arcdeg$, morphological major axis]{\includegraphics[scale=.4]{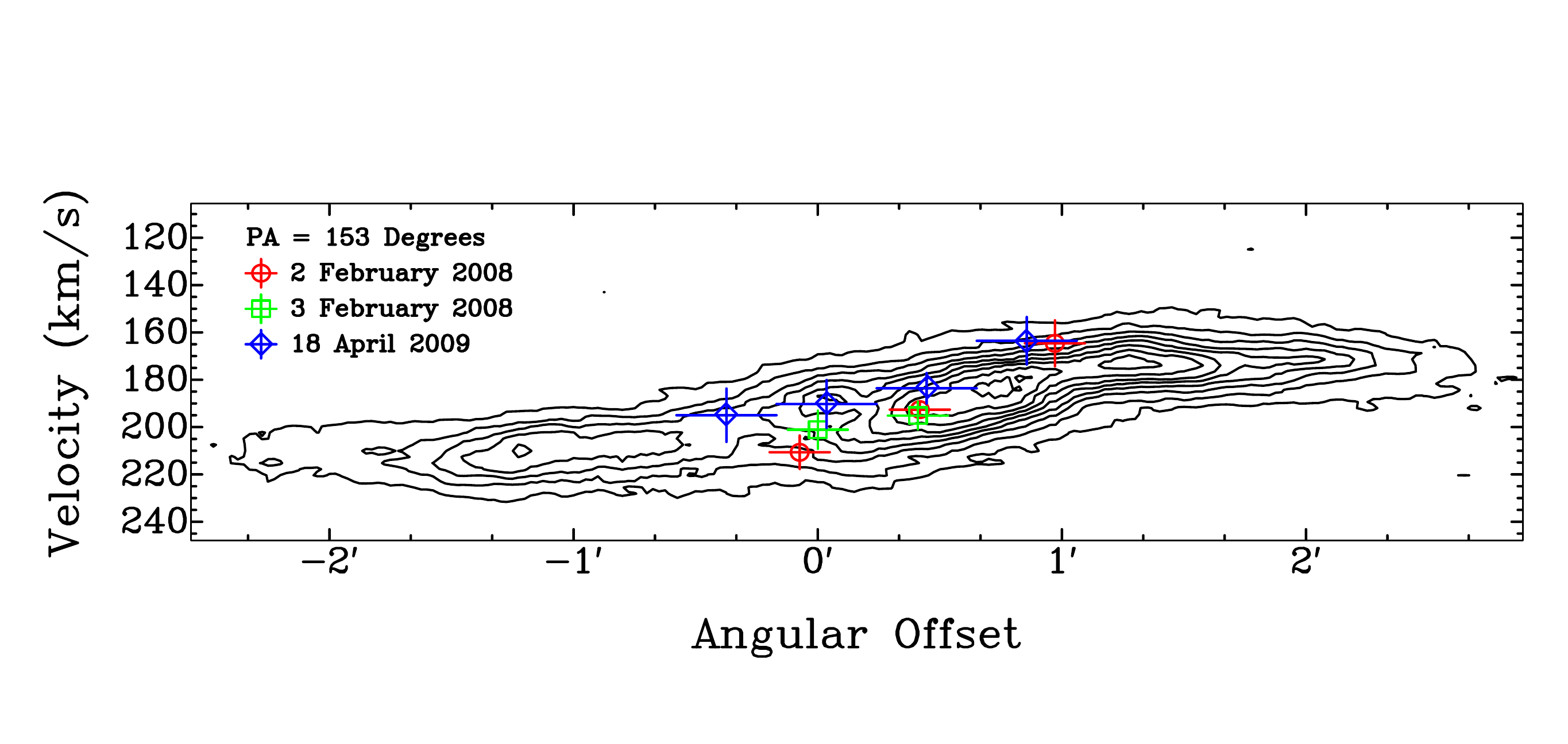}}\label{subfig:5a}
\subfigure[PA = 63$\arcdeg$, morphological minor axis]{\includegraphics[scale=.38]{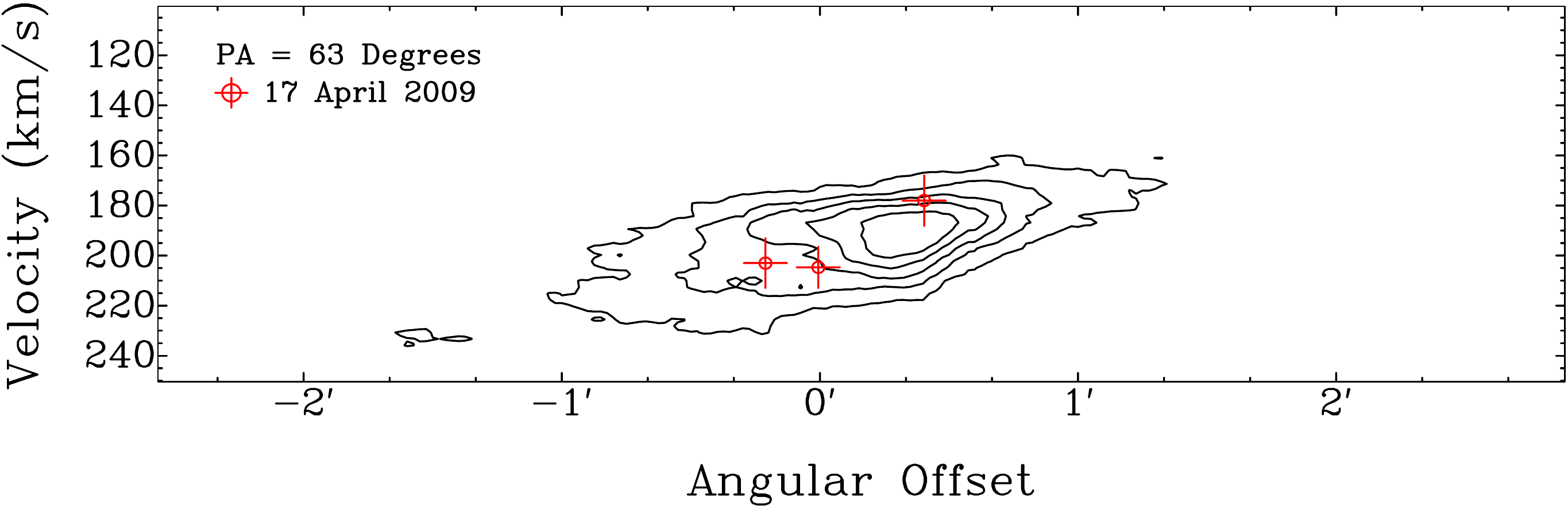}}\label{subfig:5b}
\subfigure[PA = 108$\arcdeg$, $-$45$\arcdeg$ from morphological major axis]{\includegraphics[scale=.4]{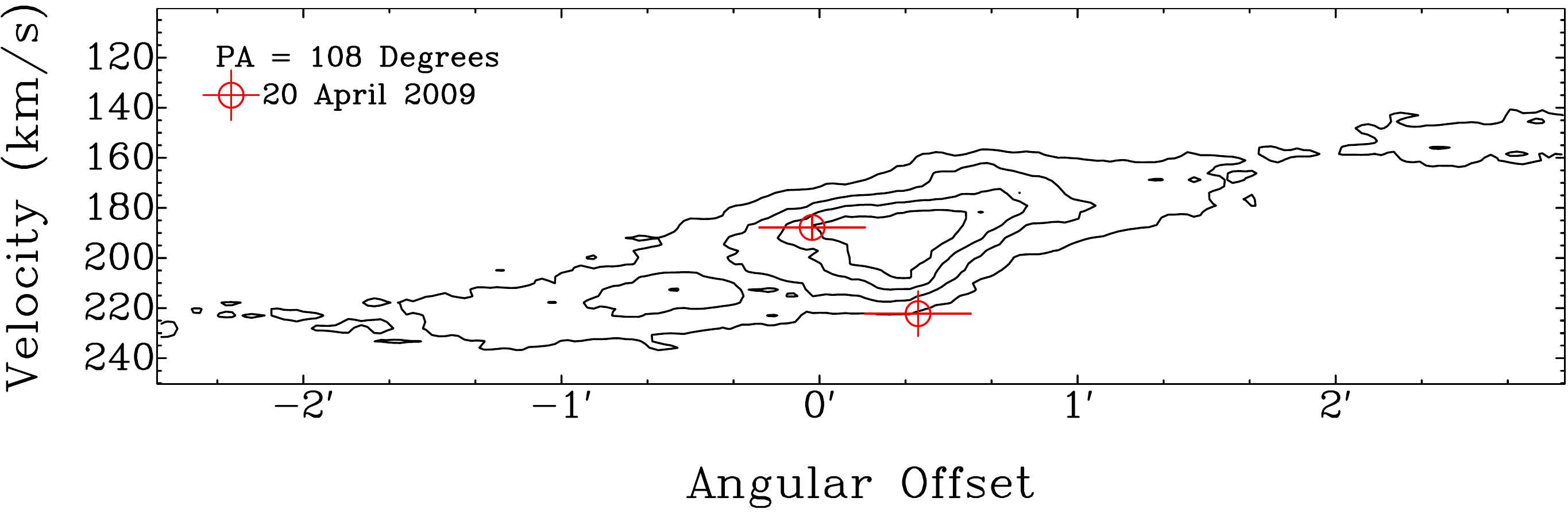}}\label{subfig:5c}
\subfigure[PA = 198$\arcdeg$, $+$45$\arcdeg$ from morphological major axis]{\includegraphics[scale=.4]{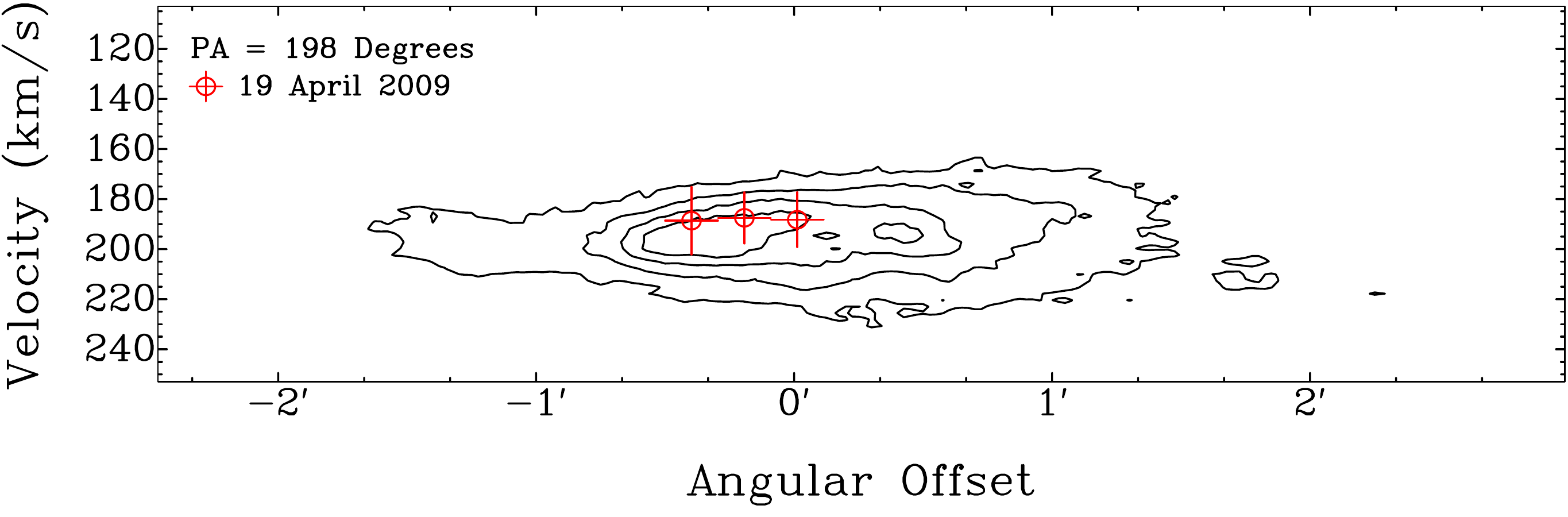}}\label{subfig:5d}
\caption{Position-velocity (P-V) diagram showing morphological major axis, PA = 153$\arcdeg$ (a), morphological minor axis, PA = 63$\arcdeg$ (b), and intermediate position angles, PA = 198$\arcdeg$ and PA = 108$\arcdeg$, (c) and (d), respectively. The intermediate position angles, PA = 198$\arcdeg$ and PA=108$\arcdeg$, are close (11$\fdg6$) to the \hi\ kinematic minor and major axes, respectively. The \hi\ is shown as contours and stellar velocity data points are plotted as colored plus symbols for DDO 168. }
\label{fig:5}
\end{figure*}

We begin our analysis by assuming that the stars in both systems follow the trends observed in spiral galaxies and have constant scale heights \citep{yoa06}, constant mass-to-light $M/L$ ratios \citep{kau03}, and are in isothermal rotating disks, the last of which is a fair assumption given that the gas disks are rotating and the stars appear to be following this rotation as shown in Figures \ref{fig:4} and \ref{fig:5}. From these assumptions, we apply the following equation from \citet{ger97}:
\begin{equation}\label{eqn:disp}
\sigma_{\rm obs}^2 = [\sigma_{\rm R}^2\ \rm sin^2(\eta - \theta)  + \sigma_{\rm \phi}^2\ cos^2(\eta - \theta )] sin^2i + \sigma_{\rm z}^2\ cos^2i
\end{equation}
where $\eta$ is the observed position angle within the plane of the galaxy, $\theta$ is the receding kinematic major axis PA and $i$ is the inclination of disk.  

Following the reasoning of \citet{hun05} and \citet{joh12}, we use the approximation from the Milky Way and other spiral galaxies that show that $\sigma_{\rm z}$/$\sigma_{\rm R}$ $\sim$ 0.7 \citep[][to within their errors]{ger97, bin98, ber10}.  
An application of the epicyclic approximation to measurements of red giant stars near the Sun where the rotation curve of the Milky Way is flat comes the relationship $\sigma_{\rm \phi}$/$\sigma_{\rm R}= 0.5[1+(d\ {\rm ln}\ v/d\ {\rm ln}\ R)]$ $\sim$ 0.7 \citep{bin93, bin98}.  These ratios, $\sigma_{\rm z}$/$\sigma_{\rm R}$ and $\sigma_{\rm \phi}$/$\sigma_{\rm R}$, appear to be constant with radius in spiral galaxies \citep{bin93, bin98, ber10} and putting these relationships together gives
$\sigma_{\rm z} \sim \sigma_{\rm \phi} \sim 0.7\sigma_{\rm R}$.  
If we take this reasoning at face value, then along the kinematic major axis position angle ($\eta - \theta$ = 0), the dependence on inclination drops out, so Equation \ref{eqn:disp} reduces to $\sigma_{\rm obs}$ = $\sigma_{\rm z}$.  

How valid the above assumptions and $\sigma_{\rm z}$/$\sigma_{\rm R}$ and $\sigma_{\rm \phi}$/$\sigma_{\rm R}$ relationships are to DDO 46 and DDO 168 is unclear.  One thing we do know, however, is that in the region where we measure the stellar velocity dispersions, the rotation curves in DDO 46 and DDO 168 are rising in a solid body manner and thus, the ratio $\sigma_{\rm \phi}/\sigma_{\rm R} = 1$ rather than 0.7.  Therefore, Equation \ref{eqn:disp} becomes 
\begin{equation}\label{eq:disper}
\sigma_{\rm z}^2 = \sigma_{\rm obs}^2 \left[\frac{\rm sin^2i}{0.49} + \rm cos^2i\right]^{-1}
\end{equation}
when $\sigma_{\rm z}$/$\sigma_{\rm R}$ $\sim$ 0.7.

For DDO 46, we suffered rainy, cloudy weather and therefore, we were only able to extract one central data point from the morphological major axis.  The velocity of this single point, 377 $\pm$ 12 \kms, matches the \hi\ gas velocity at the same location in the disk as shown in the position-velocity diagram in Figure \ref{fig:4}.  The resulting stellar velocity dispersion of this point is 15 $\pm$ 9 \kms\ and if we use the \hi\ inclination determined from the tilted-ring model of $i = 27.9$ and apply Equation \ref{eq:disper}, we obtain \sigz\ = 13.5 $\pm$ 8 \kms.

For DDO 168, the velocities of the stars match the \hi\ velocities as seen in the P-V diagrams in Figure \ref{fig:5}.  However, the morphological major axis (PA = 153$\arcdeg$) is not aligned with the kinematic major axis.  There are three pieces of evidence for this misalignment: 1) there is rotation in the stars and gas along the morphological minor axis, 2) there is a lack of rotation in both the stars and gas in PA = 198$\arcdeg$ and, 3) there is strong rotation in the \hi\ along intermediate PA = 108$\arcdeg$.  The \hi\ kinematic analysis performed using the tilted-ring model and described in Section \ref{sec:hirot}, shows that DDO 168 has a kinematic major axis for the gas of 96$\fdg4$ measured east from north in the plane of the sky (as compared to 276$\fdg4$ quoted in Table \ref{tab:1}, which is $+180\arcdeg$ from this value because the ``kinematic major axis'' is generally defined as the direction of the receding side of the galaxy).  

\begin{figure*}
\centering
\subfigure[PA = 153$\arcdeg$, morphological major axis]{\includegraphics[scale=.26]{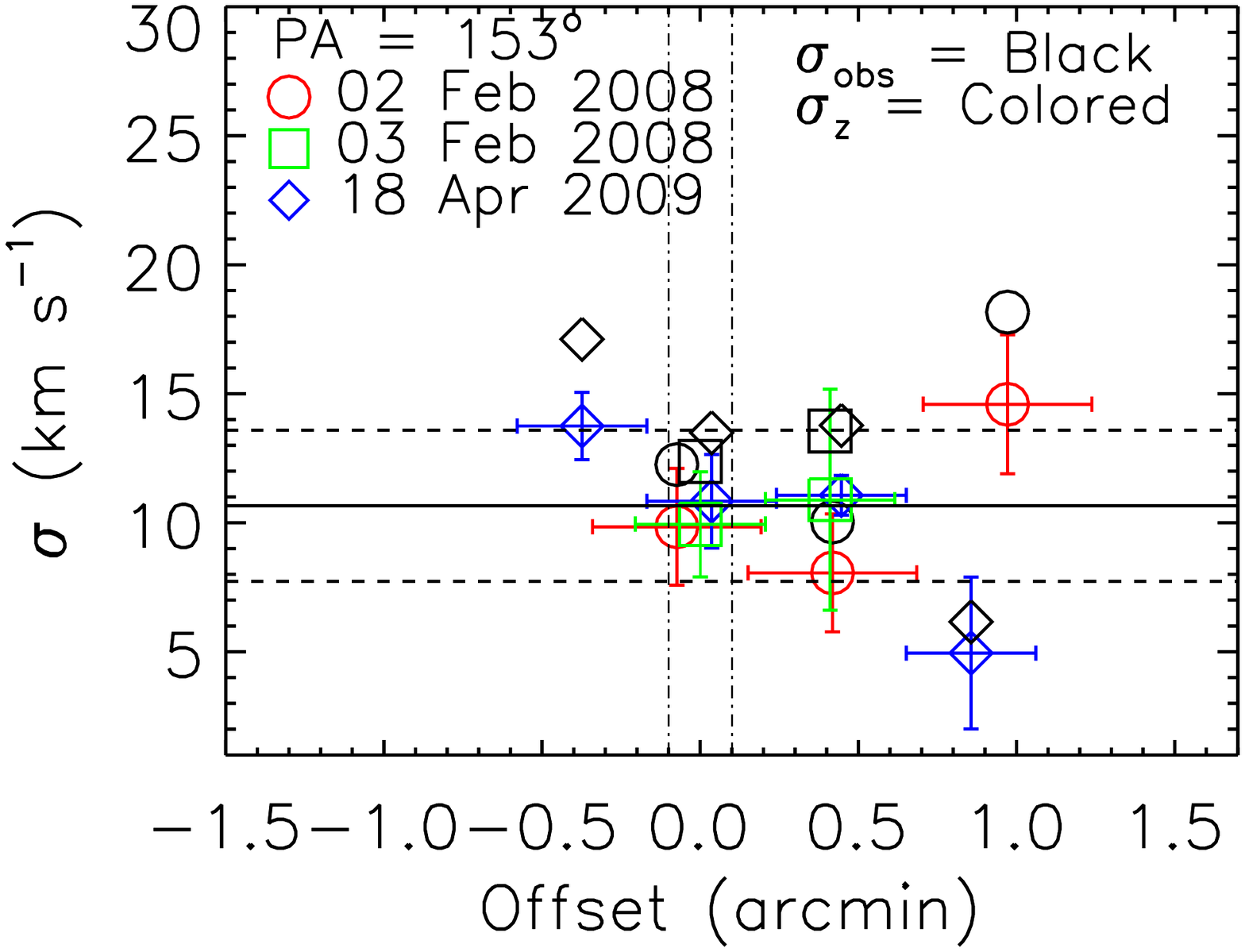}}\label{subfig:6a}
\subfigure[PA = 63$\arcdeg$, morphological minor axis]{\includegraphics[scale=.26]{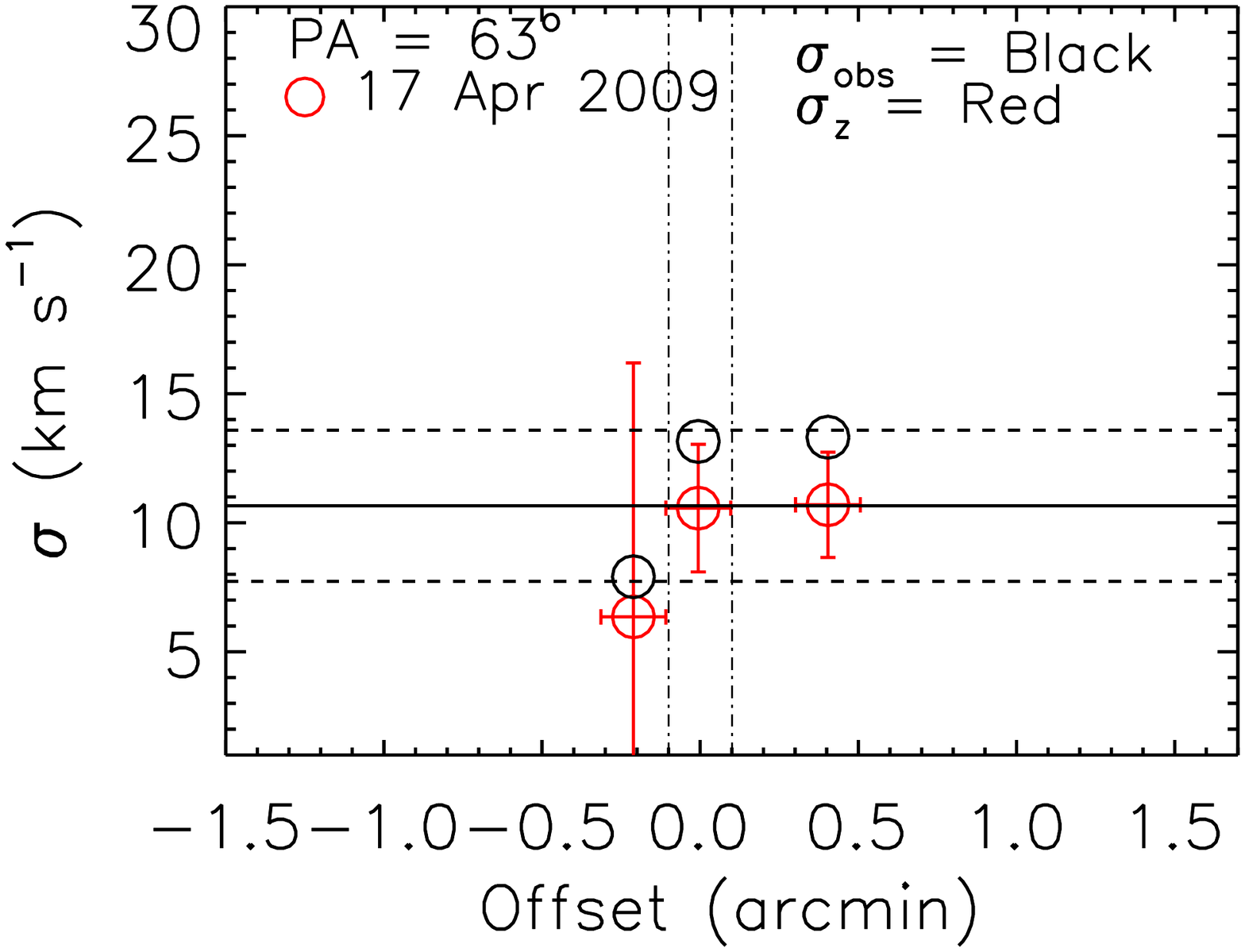}}\label{subfig:6b}
\subfigure[PA = 108$\arcdeg$, $-$45$\arcdeg$ from morphological major axis]{\includegraphics[scale=.26]{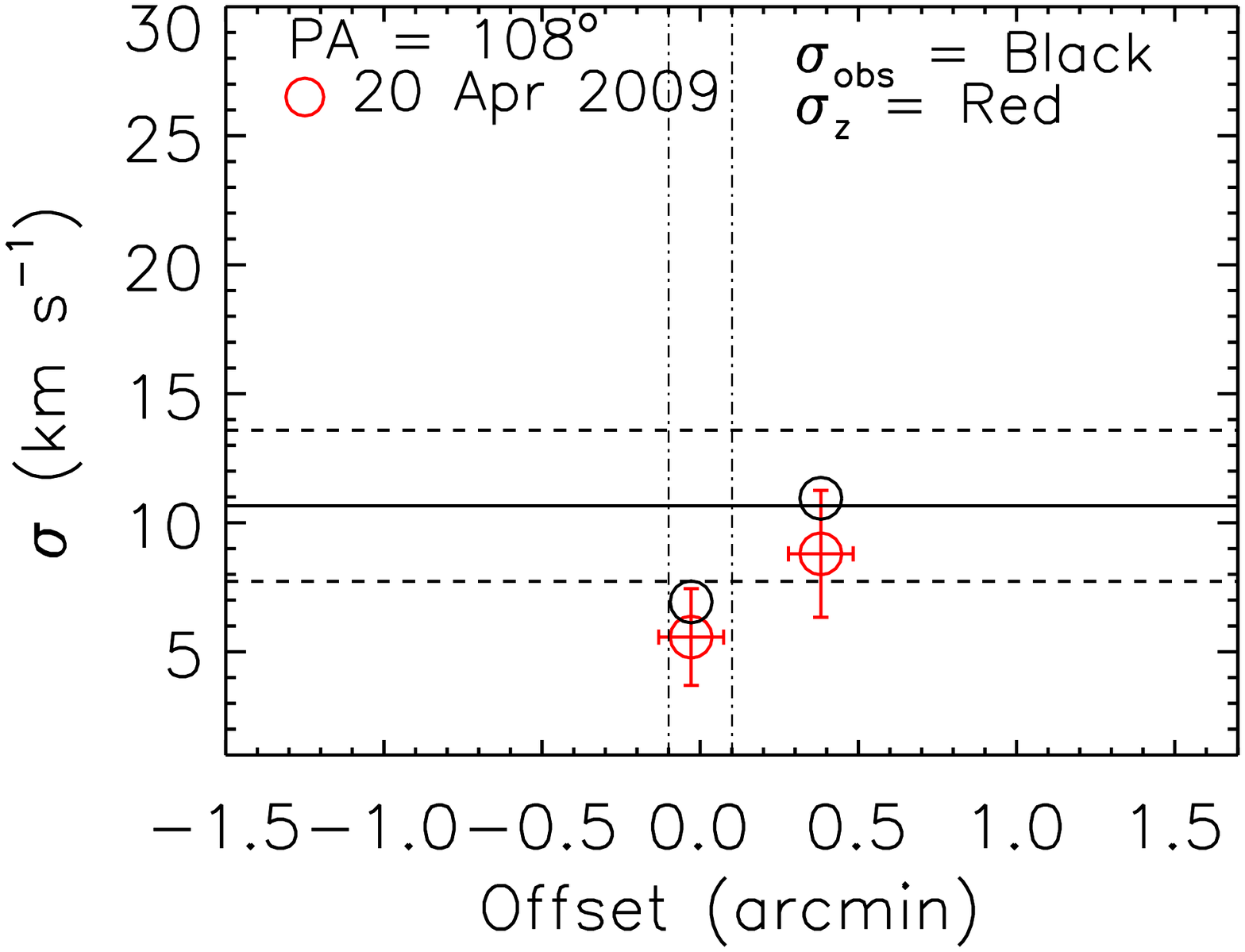}}\label{subfig:6c}
\subfigure[PA = 198$\arcdeg$, $+$45$\arcdeg$ from morphological major axis]{\includegraphics[scale=.26]{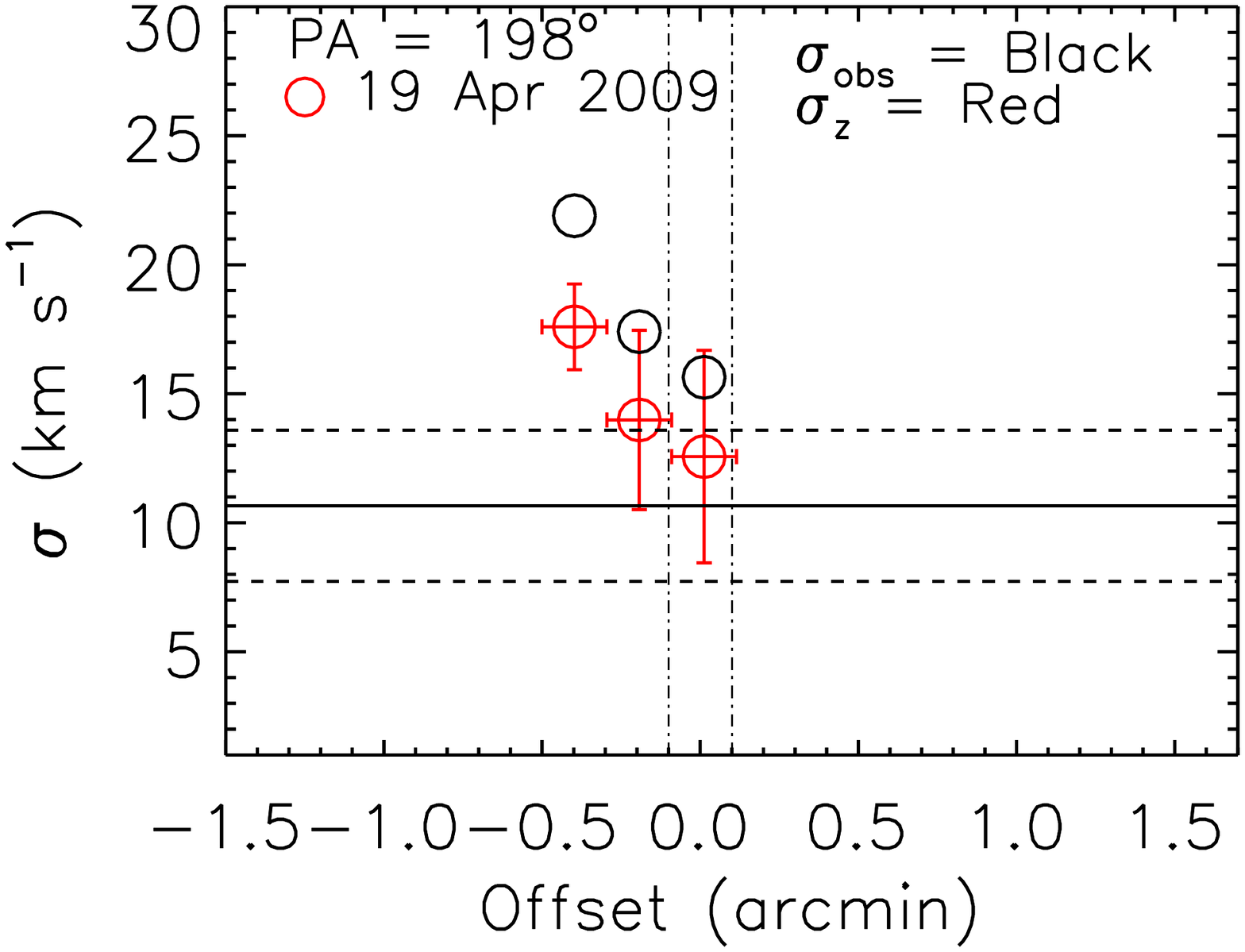}}\label{subfig:6d}
\caption{Intrinsic stellar velocity dispersions for DDO 168 determined from the CCM for all four observed position angles.  The black symbols are for the $observed$ velocity dispersions, $\sigma_{\rm obs}$, and the colored symbols are the dispersions corrected for inclination and are the disentangled $\sigma_{\rm z}$ velocities.  The horizontal solid line represents the weighted average of the central $\sigma_{\rm z}$ dispersions that lie within the vertical dot-dashed lines in all position angles. The dashed lines are the errors on this average, $<\sigma_{\rm z,0}>$ = 10.7 $\pm$ 2.9 \kms.  }
\label{fig:6}
\end{figure*}

For DDO 168,  we use inclination angle, $i$ = 46$\fdg6$ determined from the \hi\ tilted-ring analysis (described in Section \ref{sec:hirot}) and Equation \ref{eq:disper} becomes \sigz\ $ = 0.803 \sigma_{\rm obs} $ for all observed PAs.  
The results are shown in Figure \ref{fig:6}.  The colored  symbols are the $\sigma_{\rm z}$ values and the black symbols are the $\sigma_{\rm obs}$ values.  The errors were determined by calculating the standard deviation of the FWHM values from {\sc fxcor} 
using the mean of the template$-$galaxy pairs for each galaxy spectrum and then, these standard deviations were propagated through Equations \ref{eqn:sigma} (to derive the error on $\sigma_{\rm obs}$), \ref{eqn:disp}, and \ref{eq:disper} for the final errors on \sigz.
 Since the errors on $\sigma_{\rm obs}$ (black symbols) are comparable to the \sigz\ (colored symbols), they were left off Figure \ref{fig:6} for clarity.  The horizontal solid line in each of the panels of Figure \ref{fig:6} is the weighted average of all the $\sigma_{\rm z}$ values near the center, $R = 0$, and the dashed lines show the errors on this average.  The vertical dot-dashed lines show the region within which the \sigz\ values were used for determining the central velocity dispersion of DDO 168,  $<\sigma_{\rm z,0}>$ = 10.7 $\pm$ 2.9.

\section{\hi\ Spectroscopy}\label{sec:hispec}

	\subsection{Data}\label{sec:hidata}

The \hi\ data for DDO 46 and DDO 168 are from the LITTLE THINGS survey. They are VLA combined B-, C-, and D-array data that were obtained in 2008 under NRAO legacy project identification number AH927 (see Table 3 in \citet{hun12} for additional details on the observations).  The data were calibrated and imaged using the Astronomical Image Processing System (AIPS\footnote{The Astronomical Image Processing System (AIPS) has been developed by the NRAO.}).  LITTLE THINGS implemented a new multi-scale cleaning algorithm, which improves upon standard cleaning methods because the resulting \hi\ data cubes have all observed angular frequencies properly represented with near uniform noise statistics.  For more information regarding the \hi\ data calibration and imaging methods, please see \citet{hun12}.

For our purposes of comparing the stellar and gas kinematics, we use the Robust-weighted data cubes rather than the Natural-weighted cubes because the Robust-weighting produces higher spatial resolutions that are better for modeling the inner \hi\ gas kinematics where comparisons with the stars are possible. 
The Robust-weighted \hi\ data cube for DDO 46 has a beam size of 6$\farcs3$ $\times$ 5$\farcs2$, with a PA of 81$\fdg5$, while the data cube for DDO 168 has a beam size of 7$\farcs8$ $\times$ 5$\farcs8$, with a PA of 67$\fdg5$, and both have a channel spacing and spectral resolution of 2.6 \kms.

	
	\subsection{Analysis}\label{hian}
	
		\subsubsection{Double Gaussian Decomposition}\label{sec:decomp}
	
To determine the kinematics of the gas for the galaxies, we implemented a double Gaussian decomposition technique developed by \citet{oh08} in the same manner as NGC 1569 from \citet{joh12}.  This technique creates a model of the velocity field and iteratively compares the line profiles from each spatial position in the \hi\ data cubes to the model velocity field.  If there are multiple Gaussian peaks at a given location, the procedure decomposes each profile and compares the velocities of each of the peaks to the model velocity field.  The velocity that most closely matches the model is extracted into a \emph{bulk} velocity field and the outlying velocity is placed into one of two non-circular motion maps.  The \emph{strong} non-circular motion map contains the velocities whose peak intensity is \emph{higher} than the bulk velocity peak and the \emph{weak} non-circular motion map contains velocities whose peak intensity is \emph{lower} than the bulk velocity peak.  

\begin{figure*}
\centering
\includegraphics[scale=.8]{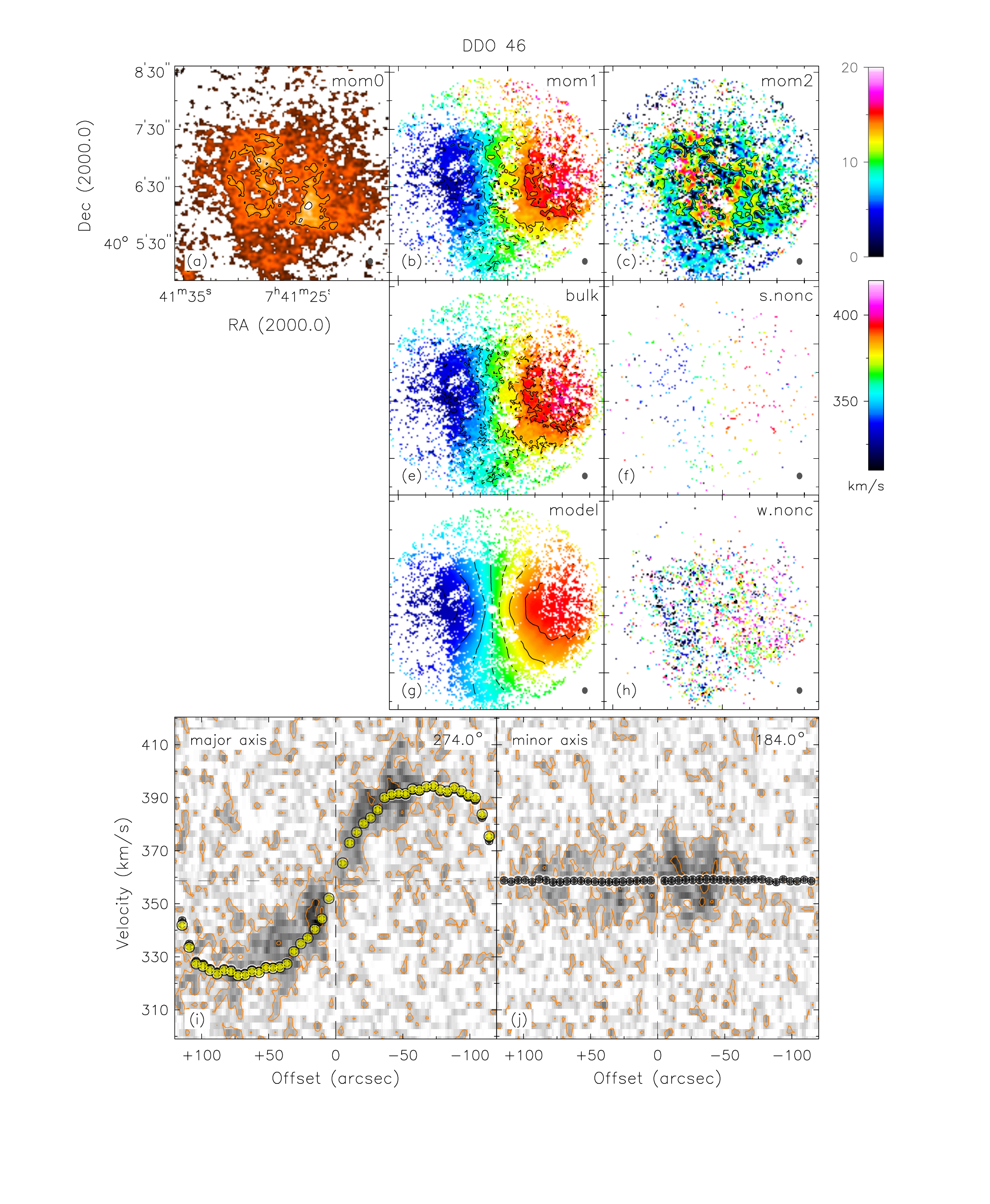}
\caption{Results of double Gaussian decomposition of the Robust-weighted \hi\ data cube for DDO 46.  Top row: integrated \hi\ intensity map; intensity-weighted \hi\ velocity field with contours separated by 10 \kms; intensity-weighted \hi\ line widths with contours separated by 5 \kms. Second row: bulk motion velocity field; strong non-circular motion map.  Third row: model velocity field used in double Gaussian decomposition; weak non-circular motion map.  Fourth row: P-V diagrams showing the bulk velocities (black points) and asymmetric drift corrected bulk velocities (yellow points) for the kinematic major (left) and minor (right) axes.  Velocity fields (b), (e), and (g) have contours 10 \kms\ apart.  See text (Section \ref{sec:decomp}) for additional explanation of individual panels.}
\label{fig:7}
\end{figure*}

\begin{figure*}
\centering
\includegraphics[scale=.8]{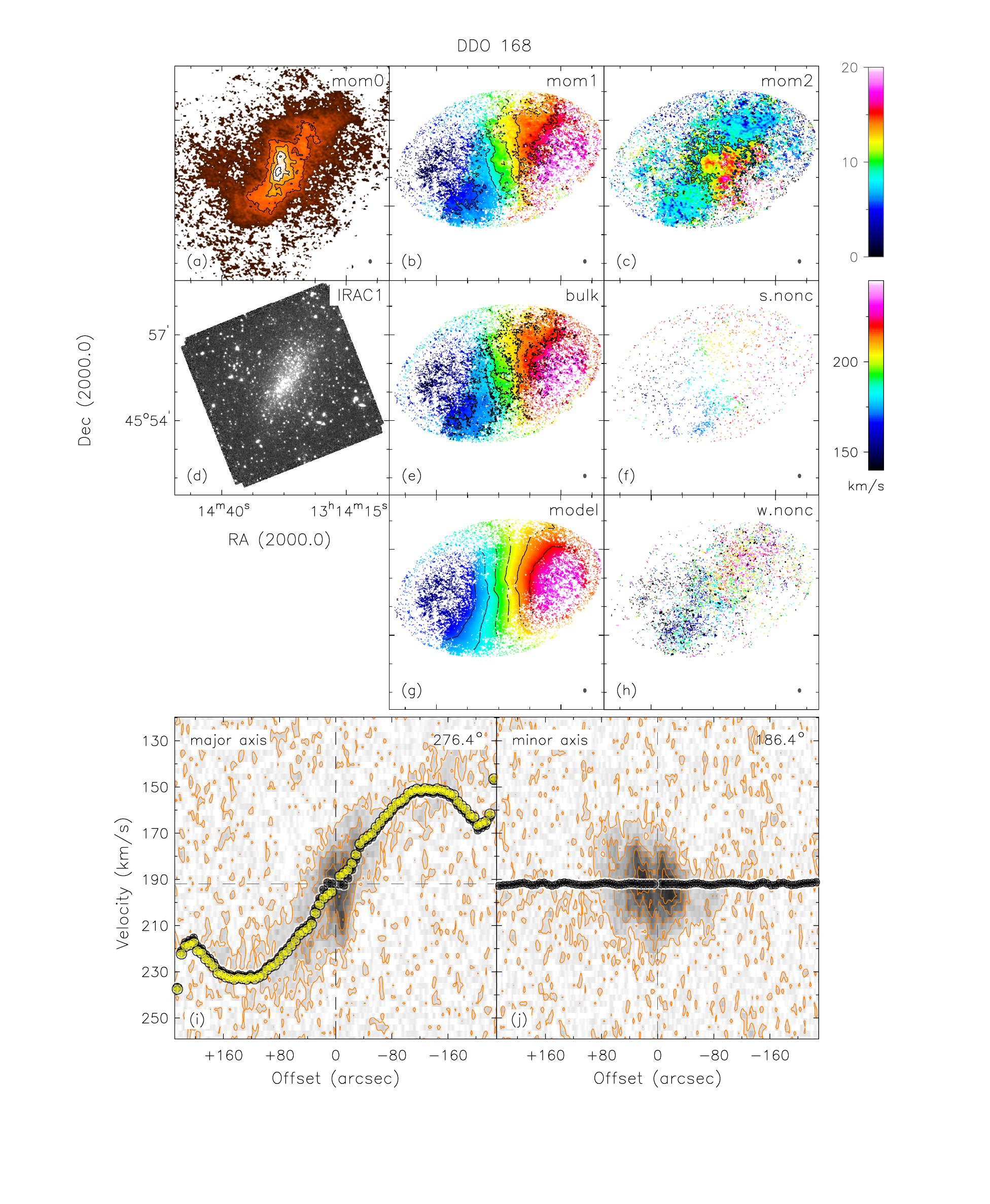}
\caption{Results of double Gaussian decomposition of the \hi\ data cube for DDO 168. Top row: integrated \hi\ intensity map; intensity-weighted \hi\ velocity field; intensity-weighted \hi\ line widths with contours separated by 5 \kms. Second row: \emph {Spitzer} 3.6$\mu$m image; bulk motion velocity field; strong non-circular motion map.  Third row: model velocity field used in double Gaussian decomposition; weak non-circular motion map.  Fourth row: P-V diagrams showing the bulk velocities (black points) and asymmetric drift corrected bulk velocities (yellow points) for the kinematic major (left) and minor (right) axes.  Velocity fields (b), (e), and (g) have contours 10 \kms\ apart.  See text (Section \ref{sec:decomp}) for additional explanation of individual panels.}
\label{fig:8}
\end{figure*}

Figures \ref{fig:7} and \ref{fig:8} show the resulting decomposed \hi\ velocity fields for DDO 46 and DDO 168, respectively.  These panels are from \citet{oh14}.  The top row in each of the figures shows the integrated intensity map, labeled ``mom0'' (0$^{\rm th}$ moment), the intensity-weighted mean velocity field, labeled ``mom1'' (1$^{\rm st}$ moment), and the intensity-weighted mean velocity dispersion map, labeled ``mom2'' (2$^{\rm nd}$ moment).  The second row in each of the figures shows the bulk velocity field, labeled ``bulk'', and the strong non-circular motion map, labeled ``s.nonc''.  In Figure \ref{fig:8} for DDO 168, the second row also contains \emph{Spitzer's} 3.6$\mu$m IRAC image for the galaxy, labeled ``IRAC''.  DDO 46 does not have IRAC 3.6$\mu$m data and is therefore, not included in Figure \ref{fig:7}.  The third row in Figures \ref{fig:7} and \ref{fig:8} show the model velocity field, labeled ``model'', and the weak non-circular motion map, labeled ``w.nonc''.  The bottom two panels in each figure show position-velocity (P-V) diagrams sliced along the kinematic major (left panel) and minor (right panel) axes as determined from the tilted-ring model.  The black data points are the bulk \hi\ rotation velocities plotted on top of the kinematic major and minor axis P-V diagrams, respectively, and the yellow data points show the bulk asymmetric drift corrected velocities along the kinematic major axis. 

		
		\subsubsection{\hi\ Rotation Curve}\label{sec:hirot}

We derived the rotation curves for DDO 46 and DDO 168 by applying the tilted-ring model to the bulk velocity field of each galaxy using the GIPSY\footnote{The Groningen Image Processing SYstem (GIPSY) has been developed by the Kapteyn Astronomical Institute.} task {\sc rotcur} \citep{beg89}.  We determined the orbital parameters in an iterative fashion for each ring independently using a ring width and spacing between rings equal to the galaxies' respective \hi\ beam widths.  The rings were fit on spatial scales from one \hi\ beam width out to a radius where the \hi\ S/N was equal to or greater than three times the rms of the noise in a line-free channel, 3$\sigma$.  The orbital parameters that were fit are: PA (position angle of the kinematic major axis), $V_{\rm sys}$ (systemic velocity of the galaxy), $i$ (inclination of the \hi\ disk), ($X_{\rm pos}, Y_{\rm pos}$) (kinematic center position), and $V_{\rm rot}$ (circular rotation velocity).  We set $V_{\rm exp}$, the expansion velocity, equal to zero.

\begin{figure*}
\centering
\includegraphics[scale=.6]{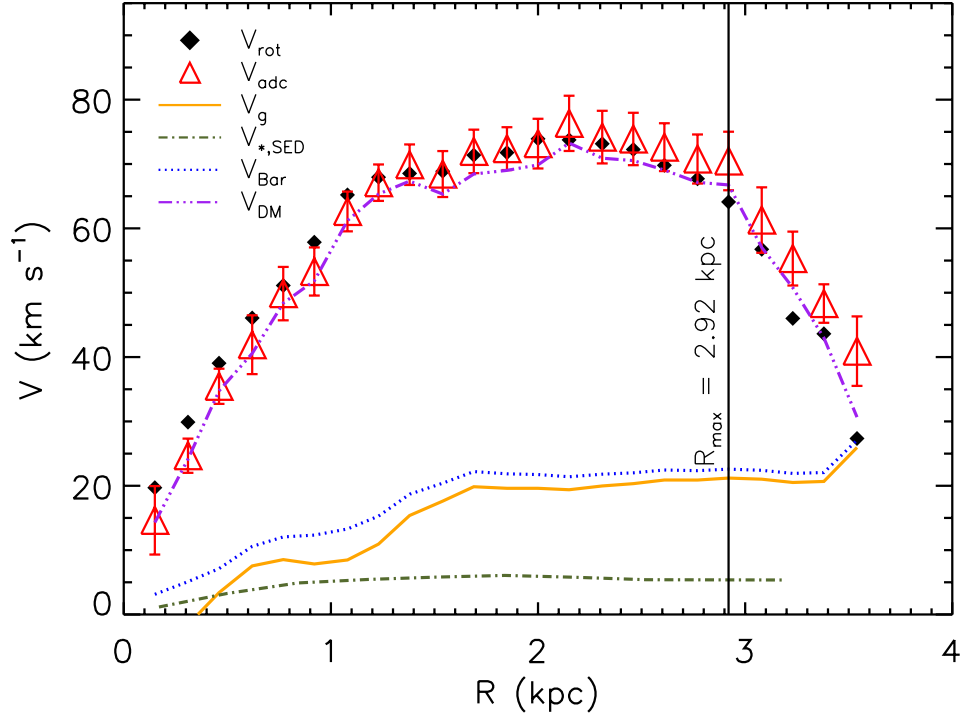}
\caption{Mass decomposition for DDO 46. The results of the tilted-ring analysis produced the rotation curve, $V_{\rm rot}$, black diamond symbols, which was then corrected for asymmetric drift, $V_{\rm adc}$, red triangle symbols.  The velocity contribution to the asymmetric drift corrected rotation curve from the mass of the gas, $V_{\rm g}$, as determined from fitting ellipses to the integrated intensity map, is shown by the orange solid line.  The velocity contribution from the stellar mass, $V_{\rm *,SED}$, as determined from the spectral energy distribution method is shown by the olive dot-dashed line.  The blue dotted line, $V_{\rm Bar}$, shows the velocity contribution to the asymmetric drift corrected rotation curve from the total baryonic mass and the purple dot-dot-dot-dashed line, $V_{\rm DM}$, shows the velocity contribution of dark matter mass. The vertical line marks the radius within which the total dynamical mass of the galaxy was determined.  Please see Section \ref{sec:massmod} for further details.}
\label{fig:9}
\end{figure*}

\begin{figure*}
\centering
\includegraphics[scale=.6]{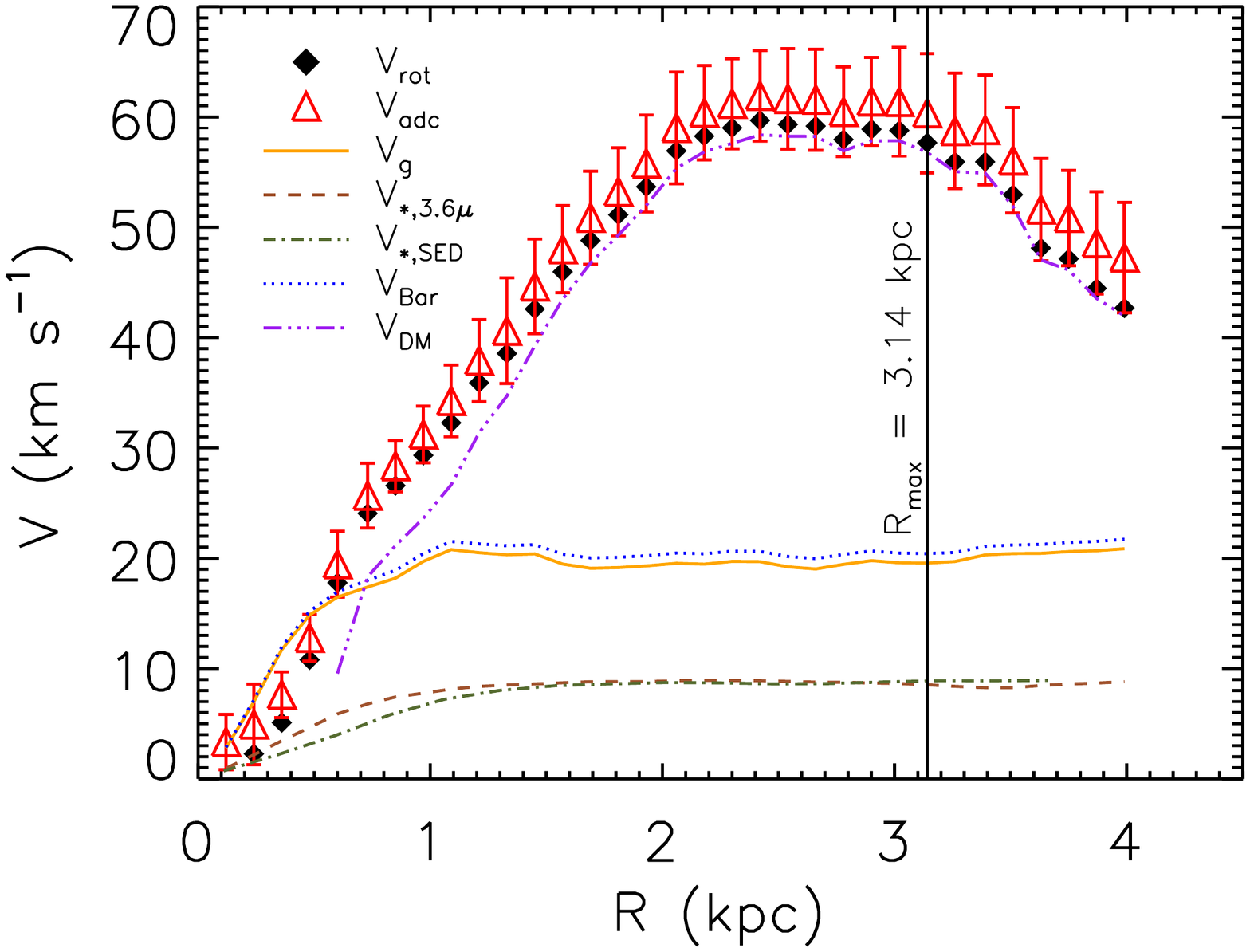}
\caption{Mass decomposition for DDO 168. The results of the tilted-ring analysis produced the rotation curve, $V_{\rm rot}$, black diamond symbols, which was then corrected for asymmetric drift, $V_{\rm adc}$, red triangle symbols.  The velocity contribution to the asymmetric drift corrected rotation curve from the mass of the gas, $V_{\rm g}$, as determined from fitting ellipses to the integrated intensity map, is shown by the orange solid line.  The velocity contribution from the stellar mass, $V_{\rm *,3.6\mu}$, as determined from ellipse fitting to the 3.6$\mu$m photometry is shown by the brown dashed line while the olive dot-dashed line shows the contribution from the stars as determined from the spectral energy distribution method, $V_{\rm *,SED}$.  The blue dotted line, $V_{\rm Bar}$, shows the velocity contribution to the asymmetric drift corrected rotation curve from the total baryonic mass and the purple dot-dot-dot-dashed line, $V_{\rm DM}$, shows the velocity contribution of dark matter mass. The vertical line marks the radius within which the total dynamical mass of the galaxy was determined.  Please see Section \ref{sec:massmod} for further details.}
\label{fig:10}
\end{figure*}

At first, we ran the tilted-ring model with all of the parameters free and we fixed one parameter at a time beginning with the kinematic center position, ($X_{\rm pos}$, $Y_{\rm pos}$).  We iterated fixing each parameter in succession and once all of the parameters were fixed except for the rotation velocity, $V_{\rm rot}$, we went through each of the variables again freeing one parameter at a time.  By iterating through the tilted-ring model six times allowing one parameter (in addition to $V_{\rm rot}$) to vary at a time, we honed in on a convergent solution for each of the orbital elements. Our results for each of the orbital elements are presented in Table \ref{tab:1} and the resulting rotation curves for DDO 46 and DDO 168 are shown by the black diamond symbols in Figures \ref{fig:9} and \ref{fig:10}, respectively. 

We correct $V_{\rm rot}$ for the asymmetric drift in the outer part of the galaxies where the dynamical support by random motions to the gas disk is significant. The asymmetric drift correction is done by the method described in \citet{bur02}:

\begin{eqnarray}
\label{eq:1}
\sigma^{2}_{\rm D} &=& -R\sigma^{2}\frac{\partial\rm ln(\rho
\sigma^{2})}{\partial R} \nonumber \\
&=& -R\sigma^{2}\frac{\partial\rm ln(\Sigma \sigma^{2})}{\partial R},
\label{sigma_D}
\end{eqnarray}

where $\sigma_{\rm D}$ is the asymmetric drift correction, $R$ is the galaxy radius,  $\sigma$ is the velocity dispersion, and $\rho$ is the volume density of gas disk. By assuming an exponential distribution in the vertical direction and a constant scale height, $\rho$ can be derived from the gas surface density $\Sigma$. The gas surface density $\Sigma$ and velocity dispersion $\sigma$ are derived by applying the tilted-rings to the integrated \hi\ (0$^{\rm th}$ moment) and velocity dispersion (2$^{\rm nd}$ moment) maps, respectively. Lastly, the asymmetric drift correction is added in quadrature to the rotation velocity from tilted-ring fits in order to derive the corrected rotation velocity $V_{\rm adc}$ as follows,

\begin{eqnarray}
\label{eq:3}
V^{2}_{\rm adc} = V^{2}_{\rm rot} + \sigma^{2}_{\rm D}.
\label{V_cor}
\end{eqnarray}

We refer to \citet{bur02} and \citet{oh11} for more details.
The resulting $V_{\rm adc}$ (with errors) are shown by the red triangle data points in Figures \ref{fig:9} and \ref{fig:10}, respectively.

\section{Results}\label{sec:results}

	\subsection{$\bf V{_{\rm \bf max}}\bf /\bf \sigma_{\rm \bf z,0}$}
	 
$V_{\rm max}/\sigma_{\rm z,0}$ for DDO 46 and DDO 168 are 5.7 $\pm$ 0.6 and 6.3 $\pm$ 0.3, respectively.  When these values are compared with those from large spiral galaxies from \citet{bot93}, they suggest that both DDO 46 and DDO 168 are kinematically cold, rotationally supported, \emph {thin} disks.  Because DDO 46 only contains one single data point for the velocity dispersion of the stars, we do not emphasize the results of the $V_{\rm max}/\sigma_{\rm z,0}$ value any further for this object.  

DDO 168, on the other hand, presents an interesting case.  As shown in Figure \ref{fig:5} and described in Section \ref{sec:optan}, the axis of rotation for this dwarf is not aligned with the morphological major or minor axes and both the gas and stars kinematically follow one another.
Together, these features suggest that DDO 168 may be kinematically heated and therefore, one might expect a $V_{\rm max}/\sigma_{\rm z,0}$ closer to one.  
When we examine the integrated \hi\ intensity map of DDO 168 shown in Figure \ref{fig:8}, we can see evidence of diffuse \hi\ emission in the outskirts of the disk.  We also observe a bar-like structure in the central regions that is seen by the \hi\ gas extending between $\sim$150 to $\sim$210 \kms\ around 0$\arcsec$ in the P-V diagram along the kinematic major axis.  \citet{bro92} observed the \hi\ distribution in DDO 168 and also found morphological peculiarities. The stars also appear to have a bar-like structure that stretches through most of the stellar disk as can be seen by the boxy $V-$band morphology in Figure \ref{fig:1}.  \citet{hun06} commented on the possible morphological misalignment between the major axis of the bar and outer disk in DDO 168.  In addition, there is kinematic evidence for a bar from the misalignment of the rotation axis of the stars and the morphological major axis. It has been shown that interactions can create tidally induced bars in disk systems \citep{miw98}, so perhaps DDO 168 is undergoing a tidal interaction from a local companion.  The most likely object that may be responsible for such an interaction is DDO 167, a dIrr galaxy that lies only 33 kpc away at the distance of DDO 168 in projection.  Despite evidence for a bar in DDO 168, there is still clear rotation in the \hi\ gas and stars and its $V_{\rm max}/\sigma_{\rm z,0}$ implies that it is a rotationally supported disk.  

If we compare the $V_{\rm max}/\sigma_{\rm z,0}$ values of DDO 46 and DDO 168 with all other dIrr galaxies that have these same measurements, we find that DDO 46 and DDO 168 are the faintest and kinematically coldest dIrr objects studied to date.  Table \ref{tab:4} lists the absolute $B$ magnitudes, stellar velocity dispersions, maximum rotation speeds and the $V_{\rm max}/\sigma_{\rm z,0}$ values for the dIrr galaxies and we include the low surface brightness Sm galaxy NGC 2552 from \citet{swa99} for comparison.  

\begin{figure*}
\centering
\includegraphics[scale=.6]{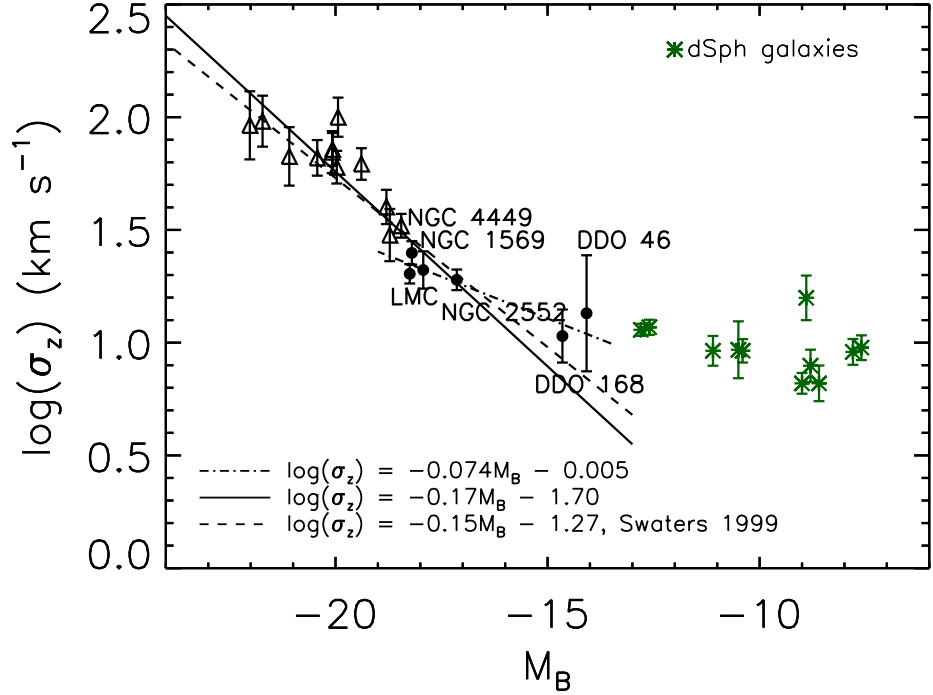}
\caption{Logarithmic stellar velocity dispersions as a function of absolute $B$ magnitudes for all dIrr objects that have measured stellar velocity dispersions (filled circles) and spiral galaxies from \citet{bot93} (triangles).  We fit a least squares line to the spiral and dIrr data, including DDO 46 and DDO 168, as shown by the solid line, and compare our fit with the one determined by \citet{swa99} shown by the dashed line. Both fits appear to agree.  We then fit a second least squares line to just the dwarfs (shown by the dot-dashed line), which has a flatter slope and may indicate a weaker (or no) trend.   Also included in the plot are the dSph galaxies for which $M_{\rm B}$ magnitudes and velocity dispersions have been measured, but, these objects are not included in any of the fits because of their innate triaxial nature.  The dSph objects are (in order of decreasing luminosity, increasing $M_{\rm B}$) Sgr, Fornax, LeoI, AndII, Sculptor, LeoII, Tucana, Sextans, Carina, Draco, and UMi; their luminosities are from \citet{mat98} and their dispersions are from \citet{wal09}.
}
\label{fig:11}
\end{figure*}

Figure \ref{fig:11} shows the logarithmic velocity dispersions of the dwarfs together with the larger spiral disk galaxies from \citet{bot93} and triaxial dSph galaxies from \citet{wal09} 
as a function of absolute $B$ magnitude, $M_{\rm B}$, in the same fashion as \citet{swa99} Figure 5, Chapter 7.  We fit a least squares line to the spiral and dIrr data, including DDO 46 and DDO 168, and determine the relationship, log($\sigma_{\rm z}$) = $-$0.17$M_{\rm B}$ $-$ 1.70.  This relationship agrees well with the relationship from \citet{swa99}.  We also fit a least squares line to just the dIrrs and find a relationship, log($\sigma_{\rm z}$) = $-$0.074$M_{\rm B}$ $-$ 0.005, which has a flatter slope possibly indicating a weaker (or even no) trend in \sigz\ with $M_{\rm B}$.  The more luminous three dIrr objects, LMC, NGC 1569 and NGC 4449, are ``puffier'' but, they are also in the throes of an interaction \citetext{LMC: \citealp{mat84, bes12}; NGC 1569: \citealp{joh13} and references therein; NGC 4449: \citealp{hun98, ric12}}.  The less luminous three, NGC 2552, DDO 168 and DDO 46, have roughly the same $\sigma_{\rm z}$ and, thus, show no obvious trend in $M_{\rm B}$ with stellar velocity dispersion.  For comparison, \citet[][see Figure 4]{wal09} find a similar flat trend in the stellar velocity dispersions of the dSphs around the Milky Way.  Although dSph galaxies are many orders of magnitude fainter and triaxial in nature and thus, are likely not a suitable comparison to the brighter, rotationally supported disk systems, it is interesting to note that as one extends to lower luminosity systems, the stellar velocity dispersions appear more constant.  More observations of dIrrs, especially in the low luminosity regime, are required to determine whether dIrrs follow the spirals, or, if they follow the more constant dispersion trend observed in the dSph systems.

Although this study is focused on the stellar kinematics, we find it pertinent to discuss the thickness of the \hi\ disk of DDO 46 and DDO 168.  Table \ref{tab:1} shows the average and central \hi\ velocity dispersions for DDO 46 and DDO 168.  It is interesting to note that the central stellar and gas dispersions for DDO 168 are nearly the same and within the errors for the single data point for DDO 46.  This is again another piece of evidence that the gas and stars in both of these galaxies are kinematically coupled.  If one determines the \hi\ \vmax/$\sigma_{\rm HI, 0}$ (using the central \hi\ velocity dispersion shown in Table \ref{tab:1}), then it appears as though DDO 46 and DDO 168 have thin \hi\ disks comparable to spiral galaxies.  However, this contradicts previous studies that have shown that \hi\ disks in dIrr systems are thicker than in spiral galaxies \citep{wal99, bri02, bag11, ban11}.  

\begin{deluxetable}{lcccc}
\tabletypesize{\scriptsize}
\tablenum{4}
\tablecolumns{5}
\tablewidth{0pt}
\tablecaption{Comparison of Stellar Kinematics in Dwarf Irregular Galaxies}
\tablehead{
\colhead{Galaxy} & \colhead{$M_{\rm B}$} &\colhead{$V_{\rm max}$} &\colhead{$\sigma_{\rm z}$} & \colhead{$V_{\rm max}/\sigma_{\rm z,0}$}\\
\colhead{Name} & &\colhead{(\kms)}&\colhead{(\kms)}&}
\startdata
LMC&$-$18.25\tablenotemark{a}&64.8 $\pm$ 15.9\tablenotemark{b}&20.2 $\pm$ 0.5\tablenotemark{b} & 2.9 $\pm$ 0.9\tablenotemark{b}\\
NGC 4449&$-$18.2\tablenotemark{c}& 80\tablenotemark{d} & 25\tablenotemark{c} & 3\tablenotemark{c}\\
NGC 1569&$-$17.93\tablenotemark{e}& 50 $\pm$ 10\tablenotemark{f} & 21 $\pm$ 4\tablenotemark{f}& 2.4 $\pm$ 0.7\tablenotemark{f}\\
NGC 2552& $-$17.5\tablenotemark{g} & 92\tablenotemark{g} & 19 $\pm$ 2\tablenotemark{g} & 5\\
DDO 168&$-$14.65\tablenotemark{d}&67.4 $\pm$ 4.0 &10.7 $\pm$ 2.9& 6.3 $\pm$ 0.3\\
DDO 46&$-$14.08\tablenotemark{d}&77.4 $\pm$ 3.7&13.5 $\pm$ 8&5.7 $\pm$ 0.6
\enddata
\tablerefs{(a) \citet{bot88};
(b) \citet{van02};
(c) \citet{hun05};
(d) \citet{hun02};
(e) \citet{hun06};
(f) \citet{joh12};
(g) \citet{swa99}}
\label{tab:4}
\end{deluxetable}

	\subsection{Mass Modeling and Dark Matter Content} \label{sec:massmod}
	In addition to the kinematic analysis for DDO 46 and DDO 168, we also decompose the observed rotation curves for each galaxy into the baryonic and dark matter mass components.  
	
		\subsubsection{Stellar Mass Model}
		
We derive the stellar contribution to the rotation curve by modeling the stellar mass surface density as a function of radius using the GIPSY task {\sc rotmod}, assuming a sech-squared disk density distribution from \citet{van81}.  The stellar mass surface density for DDO 46 and DDO 168 is found using the spectral energy distribution (SED) fitting procedure from \citet{zha12}.  We divide the Hubble time into six logarithmic age bins where we assume that the SFRs in each bin are constant.  Then, we create a library of about four million different star formation histories (SFHs) by adjusting the relative star formation intensity among the six age bins and allowing metallicities and internal dust extinction to vary. We fit the SED using Johnson's $U$, $B$, and $V$ band, \ha, and {\it Spitzer} 3.6~$\mu$m photometry data. DDO 46 does not have 3.6~$\mu$m data so this band was not used in the fit for that galaxy.  There is no remarkable global color difference between the two galaxies, except that the star formation is more intense and centrally concentrated in DDO 168. 
For more details on this SED modeling technique, see \citet{zha12}.  We correct the mass surface densities for inclination and the result of our stellar SED modeling is shown by the olive colored dot-dashed line in Figures \ref{fig:9} and \ref{fig:10} for DDO 46 and DDO 168, respectively.

Although the SED modeling is a robust method for determining the stellar mass surface density profile in a galaxy, it requires a lot of data across a large range of the spectrum from ultraviolet to infrared.  
Here, we compare the results for the stellar mass model determined from the SED modeling method with the results from using just the 3.6~$\mu$m photometry data for DDO 168 as described in \citet{oh08, oh11}.  This method bootstraps from the widely used \citet{bel01} mass-to-light ratio method for optical photometry down to the 3.6~$\mu$m wavelength.  For DDO 168, we use a ratio of scale length-to-scale height, $R_{\rm D}$/$h_{\rm z}$, of 7.3 from \citet{kre02} and a mass-to-light ratio at 3.6~$\mu$m as determined from Equation 6 in \citet{oh08}. We assume that the vertical scale height is constant across the disk and we use the isothermal relationship where 2$h_{\rm z}$ = $z_{\rm o}$ \citep{kre02, her09}.  The stellar rotation curve as determined from the stellar mass modeling using the 3.6~$\mu$m photometry data alone is shown by the dashed line in Figure \ref{fig:10}.  When compared with the stellar rotation curve derived from the SED modeling, we find that the two methods are in good agreement, especially in the outer disk.  The total stellar mass for DDO 168 is log 7.73 $M_\sun$ from the SED model and log 7.77 $M_\sun$ ($^{+0.15}_{-0.16}$ dex) from the 3.6~$\mu$m model.

		\subsubsection{Gas Mass Model and Dark Matter}
		 
We determine the mass model for the gas using the integrated \hi\ intensity maps for DDO 46 and DDO 168, shown in Figures \ref{fig:7} and \ref{fig:8}, respectively. We use the GIPSY task {\sc rotmod} and apply the \hi\ tilted-ring parameters derived from the kinematic analysis (see Section \ref{sec:hirot}) and 	
find the contribution of the gas to the rotation curve using the same assumptions as we did for the stars.  Our gas mass model takes helium and other metals into account by scaling our results by a factor of 1.36 \citep{deb08}.  The results of the gas rotation speed determined from the mass distribution of the \hi\ is shown by the orange solid line in Figures \ref{fig:9} and \ref{fig:10} for DDO 46 and DDO 168, respectively. 

To determine the total baryonic rotation curve, we add in quadrature the rotation velocities of the stars from the stellar mass model with the gas mass model velocities.  The baryonic rotation speed for DDO 46 and DDO 168 is shown by the blue dotted line in Figures \ref{fig:9} and \ref{fig:10}, respectively.  We take the baryonic rotation curve and subtract from the total observed, asymmetric drift corrected, \hi\ rotation curve and find that both DDO 46 and DDO 168 contain dark matter.  Using the maximum rotation speed from the \hi\ data, $V_{\rm max}$, we are able to determine a total dynamical mass for DDO 46 and DDO 168 using the following equation: 
\begin{equation}
M_{\rm dyn}(R) = \frac{V_{\rm max}^2\  R_{\rm max}}{G}
\end{equation}
where $R_{\rm max}$ is the radius at the last measured data point where the rotation curve is flat and $V_{\rm max}$ is measured.  From this equation, we find that DDO 46 has a dynamical mass of 4.1 $\times$ 10$^9$ $M_{\sun}$ and a dark matter mass of 3.8 $\times$ 10$^9$ $M_{\sun}$ inside $R_{\rm max}$ = 2.92 kpc (2.6$R_{\rm D}$) while DDO 168 has a dynamical mass of 3.3 $\times$ 10$^9$ $M_{\sun}$ and a dark matter mass of 3.0 $\times$ 10$^9$ $M_{\sun}$ inside $R_{\rm dyn}$ = 3.14 kpc (3.8$R_{\rm D}$).  Therefore, the global ratio of dark-to-baryonic matter mass, $M_{\rm DM}$/$M_{\rm bar}$, is $\sim$16 for DDO 46 and $\sim$9 for DDO 168.  

\begin{figure*}
\centering
\subfigure[DDO 46]{\includegraphics[scale=.4]{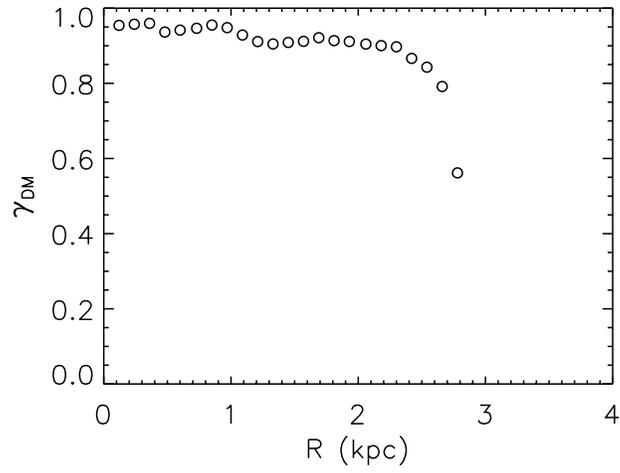}}\label{fig:12a}
\subfigure[DDO 168]{\includegraphics[scale=.4]{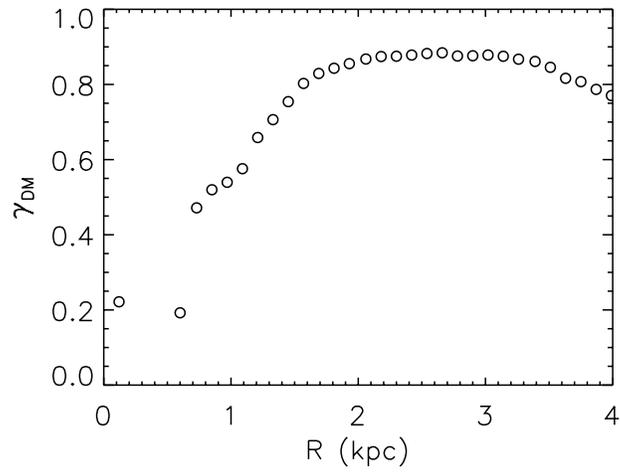}}\label{fig:12b}
\caption{The fractional dark matter content in DDO 46 (a) and DDO 168 (b). The inner $\sim$1 kpc of DDO 168 requires little to no dark matter content while DDO 46 has a dark matter fraction close to unity at nearly all radii.}
\label{fig:12}
\end{figure*}

Figure \ref{fig:12} shows the fractional dark matter content for DDO 46 and DDO 168 as a function of radius determined from the following equation:
\begin{equation}
\gamma_{\rm DM}(R) = \frac{V_{\rm adc}(R)^2 - V_{\rm gas}(R)^2 - V_{\rm *}(R)^2}{V_{\rm adc}(R)^2}
\end{equation}
where $V_{\rm adc}$ is the asymmetric drift corrected \hi\ bulk velocity, $V_{\rm gas}$ is the gas rotation velocity determined from the \hi\ surface density, and $V_{\rm *}$ is the stellar rotation velocity determined from the SED modeling described in the previous section.

In the inner $\sim$1 kpc region of DDO 168, the observed rotation curve is fit well with the total baryonic rotation speed and thus, no dark matter is required in this central region as can be seen in Figure \ref{fig:12b}.  DDO 46, on the other hand, requires dark matter at all radii and has a dark matter fraction close to one.  As previously discussed, DDO 168 is kinematically disturbed in the center and displays evidence of a bar-like structure.  DDO 168 is similar to the starburst dIrr galaxy NGC 1569 in the sense that no dark matter is required to model the observed rotation velocities in the central regions and NGC 1569 also has very disturbed kinematics \citep{joh12}.    
DDO 46 does not appear to be kinematically disturbed at any location in the disk.  Perhaps, dIrr galaxies that are kinematically disturbed in the center somehow displace their inner dark matter content.  Such a scenario has been suggested in dwarf galaxies that have high star formation rates in their centers, which can be caused by a merger or interaction event \citep{gov10}.  Both DDO 168 and NGC 1569 \citep{joh13} show signs of potential interaction or merger activity.  On the other hand, perhaps systems with little or no dark matter in the centers have less of a stabilizer against mode growth or disturbance by external perturbations.
 	
\begin{figure*}
\centering
\subfigure[DDO 46]{\includegraphics[scale=.4]{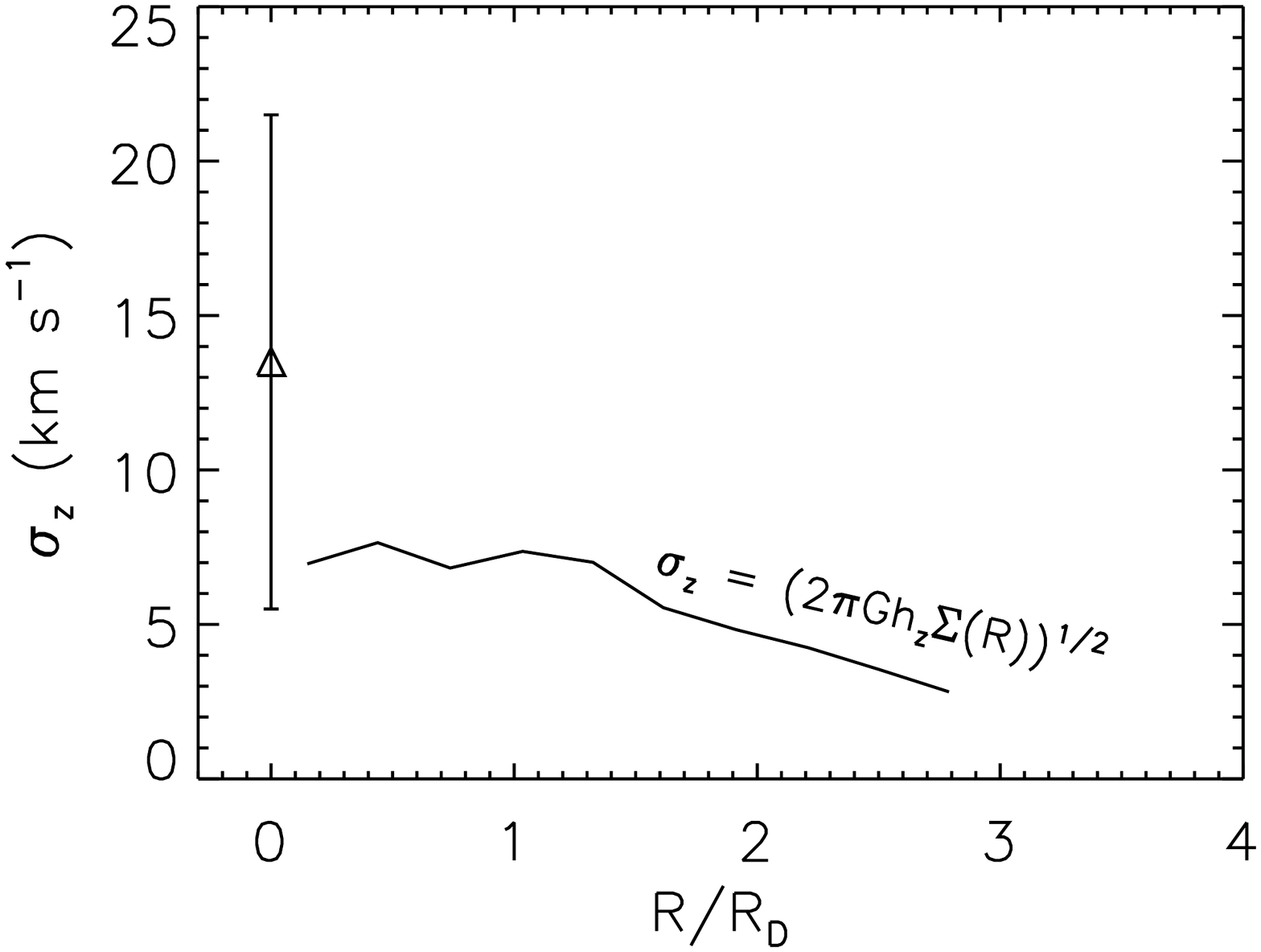}}\label{fig:13a}
\subfigure[DDO 168]{\includegraphics[scale=.4]{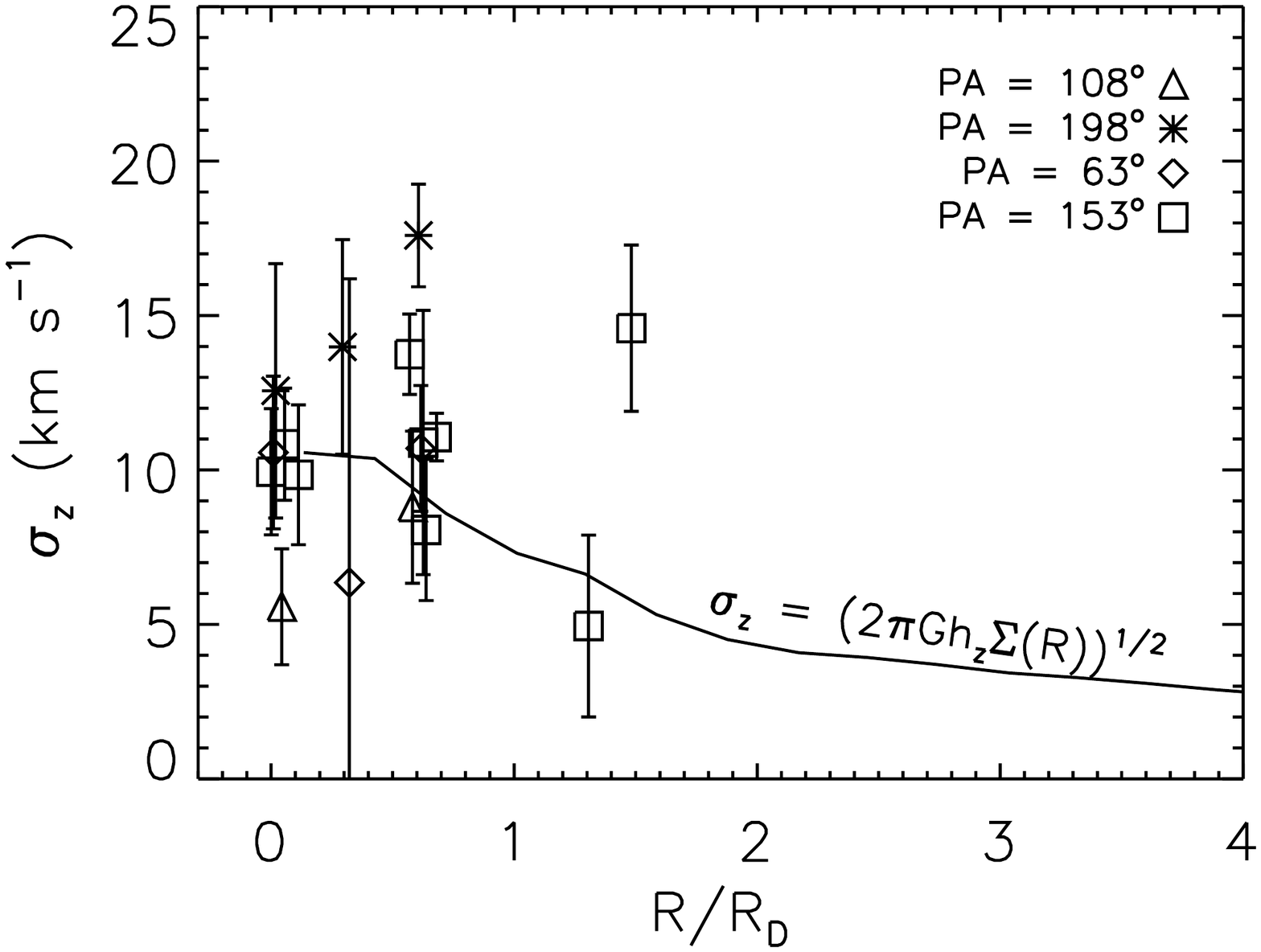}}\label{fig:13b}
\caption{The relationship between total mass surface density and stellar velocity dispersion from \citet{kre02} for (a) DDO 46 and (b) DDO 168 plotted as a function of radius in units of disk scale length, $R/R_{\rm D}$. Here, $G$ is the gravitational constant and $h_{\rm z}$ is the vertical scale height in the disk where we have assumed an isothermal distribution in the vertical direction. }\label{fig:13}
\end{figure*}

Using the total mass surface density of the stars combined with the gas, we can predict what the stellar velocity dispersion should be from the relationship derived in \citet{kre02}.  Figure \ref{fig:13} shows the relationship between stellar velocity dispersion, $\sigma_{\rm z}$, and total mass surface density $\Sigma(R)$ plotted as a function of radius in units of disk scale length, $R/R_{\rm D}$, and compares this prediction with our observed $\sigma_{\rm z}$ values.  For DDO 46, Figure \ref{fig:13} is not useful when trying to compare our observations with the prediction as we only have one data point, however, we show this figure to highlight what one might expect the velocity dispersion to be as a function of disk scale length.  Here, we expect the true $\sigma_{\rm z}$ value to be lower than what we can measure in DDO 46, thus, making \vmax/$\sigma_{\rm z,0}$ even higher.

On the other hand, it is interesting to see the comparison between the observed $\sigma_{\rm z}$ values of DDO 168 with the predicted values in Figure \ref{fig:13}(b).  There appears to be agreement between most of the predicted and observed values to within the errors, but, there are also several outliers.  A  probable explanation for these elevated dispersions may be that they are from the bar-like structure, as other studies have shown that bars can contain higher velocity dispersions than their disk hosts \citep[e.g.,][]{gad05}.

\section{Discussion}\label{sec:discuss}

This work highlights the five dIrr galaxies for which stellar velocity dispersions have been measured.  There are, however, dwarf spheroidal (dSph) galaxies around the Milky Way and M31 that have been studied kinematically \citep[e.g.,][]{wal09, tol12, ho12}.  
dSph galaxies are the lowest surface brightness, lowest mass, dark matter dominated objects in the known universe and they have virtually no \hi\ gas and are pressure supported, triaxial systems.  Because of their low surface brightness and small size, only dSph galaxies within $\sim$1 Mpc have been well studied.  The stellar kinematics for these systems are obtained by observing individual member stars and statistically deriving their global kinematic parameters.  When looking at the intrinsic stellar velocity dispersions in dSphs around the Milky Way, 
\citet{wal09} find \vmax/$\sigma$ values for eight of the Milky Way dSph galaxies and they are all $<$ 2 with an average \vmax/$\sigma$ $\sim$ 1.6.  

When looking at dSph galaxies around M31, \citet{ho12} found that the Andromeda II dSph galaxy has stellar kinematics that indicate it is rotationally supported, unlike the Milky Way dSph galaxies, and has a \vmax/$\sigma$ $>$ 1.  However, it has a dynamical mass estimate similar to the dynamical masses of the Milky Way dSphs from \citet{wal09}.  \citet{tol12} studied the stellar kinematics of 15 dSph galaxies around M31 and concluded that the kinematics of the objects in their sample are fully consistent with the results of the Milky Way dSph galaxies.  
Comparing the stellar kinematics of dSphs with dIrrs, we find that dSphs show evidence for a pressure supported triaxial figure with perhaps a small amount of rotation, as their  \vmax/$\sigma$ $\sim$ 1 values suggest, while dIrrs are disk structures that are dominated by rotation.  

If we compare the \vmax/$\sigma_{\rm z,0}$ values with those of spiral galaxies \citep{bot93, veg01}, we find that the dIrrs appear to have \emph {thin} disks.  This is in contrast with the studies of minor-to-major photometric axis ratio ($b/a$) studies, which suggest that dIrrs are \emph{thick} disks \citep{hod66, van88, sta92, roy13}.
If dIrr galaxies are thin, how can we explain the distribution of $b/a$ axis ratios?  
Perhaps dIrrs are not circular, as all of these studies, including this one, always assume, but rather, they are elliptical.  Perhaps, \vmax/$\sigma_{\rm z,0}$\ in these systems is not indicative of the global shape of dIrr galaxies.  In DDO 168, we measure \vmax/$\sigma_{\rm z,0}$\ where \sigz\ is measured only in the central regions.  At $\sim$1 kpc, $V_{\rm rot}$ is roughly two times lower than \vmax\ and this is the radius where \sigz\ becomes too faint to measure, so, \vmax/$\sigma_{\rm z}$\ locally may be two times lower.  It's not clear, however, what a local $V_{\rm rot}$/\sigz\ value means in these systems.  If \sigz\ drops off with radius as Figure \ref{fig:13}(b) suggests, then \vmax/$\sigma_{\rm z}$ rises with increasing radius, thus, dIrrs may be fat in the center and thin in the outer regions. Here again, more stellar kinematic observations of dIrr galaxies are required to understand their innate three dimensional shape.


\section{Summary \& Conclusions}\label{sec:sumcon}

	\subsection{Stellar Kinematics}
	
\noindent 1) We find a central stellar velocity dispersion perpendicular to the disk, $\sigma_{\rm z,0}$, of 13.5 $\pm$ 8 \kms\ and 10.7 $\pm$ 2.9 \kms\ for DDO 46 and DDO 168, respectively.  The stellar dispersion for DDO 46 was measured from a single central data point while DDO 168's dispersion is the average of the center positions observed in four PAs, corrected for inclination, assuming a disk geometry similar to the Milky Way.\\
2) The axis of rotation for DDO 168 is misaligned with respect to the morphological major axis and the stars show morphological and kinematic evidence of a bar structure.\\
3) The stars and gas in DDO 168 appear to kinematically follow one another.
	
	\subsection{\hi\ Kinematics}
	
\noindent 1) The \hi\ bulk velocity fields of DDO 46 and DDO 168 show rotation.  The axis of rotation is aligned with the morphological major axis in DDO 46  but it is misaligned in DDO 168. \\
2) We determine a maximum rotation velocity, corrected for asymmetric drift, \vmax, of 77.4 $\pm$ 3.7 \kms\ for DDO 46 and 67.4 $\pm$ 4.0 \kms\ for DDO 168.  These values combine with the \sigz\ values from the stars to produce kinematic measures, \vmax/$\sigma_{\rm z,0}$, of 5.7 $\pm$ 0.6 for DDO 46 and 6.3 $\pm$ 0.3 for DDO 168, which are both indicative of $thin$ disks.\\
3) There is little emission in the non-circular motion maps for both DDO 46 and DDO 168.  However, the \hi\ morphology and kinematics of DDO 168 show signs of a bar that stretches over much of the disk.\\
4) There is tenuous, extended emission in the outskirts of DDO 168 that hints at a potential interaction or recent merger.
 
	\subsection{Mass Modeling}
	
\noindent 1) We determine a total stellar mass of 2.45 $\times$ 10$^7$ \msun\ for DDO 46 and 5.90 $\times$ 10$^7$ \msun\ for DDO 168 by modeling the SED using infrared, optical, and ultraviolet photometry data.\\
2) We find a total gas mass of 2.7 $\times$ 10$^8$ \msun\ for DDO 46 and 3.5 $\times$ 10$^8$ \msun\ for DDO 168.\\
3) We derive a total dynamical mass of 3.6 $\times$ 10$^9$ \msun\ for DDO 46 and 2.2 $\times$ 10$^9$ \msun\ for DDO 168, which leads to a total dark matter content of 3.3 $\times$ 10$^9$ \msun\ for DDO 46 and 1.8 $\times$ 10$^9$ \msun\ for DDO 168.\\
4) The baryons dominate the inner $\sim$1 kpc region of the observed rotation curve in DDO 168 and thus, no dark matter is necessary in this central area.  Conversely, dark matter is required at all radii in DDO 46 in order to model the observed rotation curve.\\
5) We find a global dark-to-baryonic matter mass ratio, $M_{\rm DM} / M_{\rm bar}$, of 11 for DDO 46 and 4 for DDO 168.
	
	\subsection{Comparison with other disk galaxies}
	
DDO 46 and DDO 168 are two out of five dIrr systems for which stellar velocity dispersions have been measured.  From these five systems, there does not appear to be a trend in luminosity with shape.  On the other hand, we find that DDO 46 and DDO 168 agree with the \sigz\ $-$ $M_{\rm B}$ relationship from \citet{bot93} and \citet{swa99}, although the errors on our \sigz\ values are large.  When compared with dSph galaxies around the Milky Way and M31, we see that dSphs are triaxial, pressure supported objects while dIrr systems are rotationally supported disk systems, although we note that dSph galaxies are several orders of magnitude fainter than the dIrr systems studied to date, as shown in Figure \ref{fig:11}.  
	

\section{Acknowledgments}

The authors are grateful to the referee for their useful comments and suggestions. This project was funded by the National Science Foundation under the LITTLE THINGS grant number AST-0707563 to DAH.  This research has made use of the NASA/IPAC Extragalactic Database (NED) which is operated by the Jet Propulsion Laboratory, California Institute of Technology, under contract with the National Aeronautics and Space Administration.



\clearpage

\begin{deluxetable}{cccccccccc}
\tabletypesize{\tiny}
\tablecaption{Radial velocity standard stars used in cross-correlation method.\tablenotemark{*}}
\tablenum{3}
\tablecolumns{11}
\tablewidth{0pt}
\tablehead{\colhead{HD} & & \colhead{Metallicity} & \colhead{Spectral} & \colhead{RA (J2000)} & \colhead{DEC (J2000)} & \colhead{$V_{\rm helio}$\tablenotemark{a}} 
& \colhead{$T_{\rm exp}$} & \colhead{DDO 46} & \colhead{DDO 168}\\ 
\colhead{No.} & \colhead{$V$} & \colhead{[Fe/H]} & \colhead{Type \& Class} & \colhead{hh:mm:ss} & \colhead{$\arcdeg$ : $\arcmin$ : $\arcsec$} & \colhead{\kms} 
&\colhead{(sec)} &\colhead{UT Date Obs.}&\colhead{UT Date Obs.}}
\startdata
3765&7.36&0.10\tablenotemark{1}&K2 V&00:41:17.4&+40:13:56&-63.0 $\pm$ 0.2&90&16 Jan 10&\\
4388&7.34&\nodata&K3 III&00:46:54.5&+30:59:52&-28.3 $\pm$ 0.6&70&15 Jan 10\\
8779&6.41&-0.40\tablenotemark{2}&K0 IV&01:26:53.5&-00:21:18&-5.0 $\pm$ 0.5&36&&03 Feb 08\\
	&&&&&&&45&16 Jan 10&\\
	&&&&&&&45&17 Jan 10&\\
9138 &4.84&-0.37\tablenotemark{3}&K4 III&01:30:37.9&+06:11:15&+35.4 $\pm$ 0.5&7&&03 Feb 08\\
	&&&&&&&15&16 Jan 10&\\
	&&&&&&&15&17 Jan 10&\\
12029&7.44&\nodata&K2 III&01:59:11.2&+29:25:16&+38.6 $\pm$ 0.5&80&16 Jan 10&\\
	&&&&&&&80&17 Jan 10&\\
22484 &4.28&-0.10\tablenotemark{3}&F9 IV-V&03:37:18.5&+00:25:41&+27.9 $\pm$ 0.1&5 &17 Jan 10&02 Feb 08\\
	&&&&&&&5&&03 Feb 08\\
	&&&&&&&10&&\\
23169&8.50&\nodata&G2 V&03:44:23.9&+25:45:06&+13.3 $\pm$ 0.2&240&&03 Feb 08\\
	&&&&&&&170&15 Jan 10&\\
	&&&&&&&170&17 Jan 10&\\
	&&&&&&&220&&\\
26162 & 5.50 & -0.02\tablenotemark{1}&K1 III & 04:09:39.8 & +19:37:52 & +23.9 $\pm$ 0.6 & 30&18 Jan 10&\\
32963&7.60&0.08\tablenotemark{1}&G5 IV& 05:08:27.3&+26:20:18&-63.1 $\pm$ 0.4&120&&02 Feb 08\\
	&&&&&&&120&&19 Apr 09\\
	&&&&&&&90&15 Jan 10&\\
	&&&&&&&90&16 Jan 10&\\
	&&&&&&&90&17 Jan 10&\\
	&&&&&&&90&18 Jan 10&\\
	&&&&&&&100&&\\
65583&6.97&-0.56\tablenotemark{3}&G8 V&08:01:03.6&+29:11:09&+12.5 $\pm$ 0.4&60&&19 Apr 09\\
	&&&&&&&80&&\\
65934&7.70&\nodata&G8 III&08:02:42.1&+26:36:49&+35.0 $\pm$ 0.3&120&&20 Apr 09\\
	&&&&&&&100&&\\
66141&4.39&-0.30\tablenotemark{3}&K2 IIIb Fe-0.5&08:02:42.5&+02:18:39&+70.9 $\pm$ 0.3&15&&18 Apr 09\\
	&&&&&&&10&&19 Apr 09\\
	&&&&&&&30&&20 Apr 09\\
75935&8.46&\nodata&G8 V&08:54:20.3&+26:52:50&-18.9 $\pm$ 0.3&170&16 Jan 10&17 Apr 09\\
	&&&&&&&240&17 Jan 10&\\
	&&&&&&&200&&\\
90861&6.88&\nodata&K2 III&10:30:22.2&+28:32:15&+36.3 $\pm$ 0.4&60&16 Jan 10&17 Apr 09\\
	&&&&&&&90&17 Jan 10&20 Apr 09\\
	&&&&&&&80&&\\
	&&&&&&&70&&\\
92588&6.26&-0.10\tablenotemark{2}&K1 IV&10:41:50.1&-01:47:11&+42.8 $\pm$ 0.1&35&16 Jan 10&18 Apr 09\\
	&&&&&&&35&&20 Apr 09\\
	&&&&&&&60&17 Jan 10&\\
102494&7.48&-0.26\tablenotemark{4}&G9 IVw...&11:48:22.8&+27:17:36&-22.9 $\pm$ 0.3&90&16 Jan 10&19 Apr 09\\
	&&&&&&&100&&20 Apr 09\\
	&&&&&&&300&17 Jan 10&\\
103095&6.45&-1.12\tablenotemark{9}&G8 Vp&11:53:28.0&+37:39:28&-99.1 $\pm$ 0.3&50&&20 Apr 09\\
	&&&&&&&90&17 Jan 10&\\
107328&4.96&-0.46\tablenotemark{8}&K0.5 IIIb Fe-0.5&12:20:46.9&+03:15:55&35.7 $\pm$ 0.3&10&&19 Apr 09\\
	&&&&&&&12&&20 Apr 09\\
122693&8.11&\nodata&F8 V&14:03:15.6&+24:31:14&-6.3 $\pm$ 0.2&170&18 Jan 10&03 Feb 08\\
	&&&&&&&80&&20 Apr 09\\
	&&&&&&&200&&\\
126053&6.27&-0.45\tablenotemark{3}&G1 V&14:23:41.4&+01:12:08&-18.5 $\pm$ 0.4&40&18 Jan 10&02 Feb 08\\
	&&&&&&&35&&03 Feb 08\\
	&&&&&&&40&&17 Apr 09\\
	&&&&&&&40&&20 Apr 09\\
132737&7.64&\nodata&K0 III&15:00:14.4&+27:07:37&-24.1 $\pm$ 0.3&80&&02 Feb 08\\
	&&&&&&&80&&03 Feb 08\\
	&&&&&&&80&&18 Apr 09\\
	&&&&&&&200&18 Jan 10&\\
136202&5.06&-0.08\tablenotemark{3}&F8 III-IV&15:19:44.9&+01:44:01&+53.5 $\pm$ 0.2&24&&03 Feb 08\\
	&&&&&&&24&&17 Apr 09\\
	&&&&&&&30&&\\
144579 &6.66&-0.69\tablenotemark{1}&G8 IV&16:05:14.4&+39:08:02&-60.0 $\pm$ 0.3&35&&03 Feb 08\\
	&&&&&&&35&&19 Apr 09\\
	&&&&&&&50&&\\
145001&5.00&-0.26\tablenotemark{5}&G5 III&16:08:27.6&+17:01:29&-9.5 $\pm$ 0.2&15&&02 Feb 08\\
	&&&&&&&10&&03 Feb 08\\
	&&&&&&&10&&18 Apr 09\\
	&&&&&&&20&&19 Apr 09\\
	&&&&&&&10&&20 Apr 09\\
	&&&&&&&25&&\\
154417&6.01&-0.04\tablenotemark{6}&F8.5 IV-V&17:05:42.8&+00:41:26&-17.4 $\pm$ 0.3&20&&03 Feb 08\\
	&&&&&&&35&&17 Apr 09\\
182572&5.16&-0.37\tablenotemark{1}&G7 IV H$\delta$ 1&19:25:22.5&+11:57:47&-100.5 $\pm$ 0.4&24&&17 Apr 09\\
	&&&&&&&24&&18 Apr 09\\
	&&&&&&&24&&19 Apr 09\\
	&&&&&&&40&&20 Apr 09\\
187691&5.11&0.07\tablenotemark{7}&F8 V&19:51:26.1&+10:26:15&+0.1 $\pm$ 0.3&24&&18 Apr 09\\
212493&4.79&\nodata&K0 III&22:28:17.3&+04:44:19&+54.3 $\pm$ 0.3&20&18 Jan 10\\
213947&6.88&\nodata&K2&22:35:00.6&+26:38:32&+16.7 $\pm$ 0.3&45&15 Jan 10&\\
	&&&&&&&45&18 Jan 10&\\
222368&4.13&-0.15\tablenotemark{3}&F7 V&23:40:23.3&+05:40:21&+5.3 $\pm$ 0.2&5&15 Jan 10\\
	&&&&&&&5&16 Jan 10&\\
	&&&&&&&8&17 Jan 10&\\	
	&&&&&&&5&18 Jan 10&\\
223311&6.07&\nodata&K4 III&23:48:58.7&-06:20:00&-20.4 $\pm$ 0.1&30&15 Jan 10&\\
	&&&&&&&30&16 Jan 10&\\
	&&&&&&&30&17 Jan 10&\\
Sun\tablenotemark{b}&4.83&0&G2 V&\nodata&\nodata&\nodata&\nodata&\nodata&02 Feb 08\\
	&&&&&&&&&03 Feb 08\\
	&&&&&&&&&18 Apr 09\\
	&&&&&&&&&19 Apr 09\\
	&&&&&&&&&20 Apr 09\\
\enddata	
\tablenotetext{*}{All data in table, except for metallicity and the last two columns, come from the United States Nautical Almanac (2008).  Metallicities are listed, where available, and references are given below.}
\tablenotetext{a}{Heliocentric radial velocities.}
\tablenotetext{b}{We extracted a one-dimensional spectrum from the combined twilight flats on the nights we obtained them and used the spectrum as a standard star template in the cross-correlation method, as explained in Section \ref{sec:optobs}.}
\tablerefs{(1) \citet{sou08};
(2) \citet{ran99};
(3) \citet{cen07}; 
(4) \citet{yos97}; 
(5) \citet{mcw90}; 
(6) \citet{che00}; 
(7) \citet{fuh98}; 
(8) \citet{luc07}; 
(9) \citet{mal12}}

\label{tab:3}
\end{deluxetable}	

\clearpage

\end{document}